\documentclass[longauth]{aa}

\usepackage{natbib}
\bibpunct{(}{)}{;}{a}{}{,} 
\usepackage{amssymb}
\usepackage{url}
\usepackage{hyperref}
\usepackage{graphicx}
\usepackage{longtable}
\usepackage{booktabs}
\usepackage{lscape}
\newcommand{\urltt}[1]{\texttt{#1}}
\newcommand{\micron}{\mbox{$\mu$m}}
\newcommand{\msun}{\mbox{$M_\odot$}}
\newcommand{\mstar}{\mbox{$M_*$}}
\newcommand{\lsun}{\mbox{$L_\odot$}}
\newcommand{\msunyr}{\mbox{\msun\ yr$^{-1}$}}
\newcommand{\hi}{\ion{H}{i}}
\newcommand{\mhi}{\mbox{$M_{\rm H{\sc I}}$}}
\newcommand{\zhi}{\mbox{$z_{\rm HI}$}}
\newcommand{\lphi}{\mbox{$L'_{\rm HI}$}}

\newcommand{\lp}{\mbox{$L'$}}
\newcommand{\kms}{\mbox{km\,s$^{-1}$}}
\newcommand{\Kkmspc}{\mbox{K\,km\,s$^{-1}$\,pc$^2$}}
\newcommand{\htwo}{\mbox{H$_2$}}
\newcommand{\mhtwo}{\mbox{$M_{\rm H_2}$}}
\newcommand{\metoh}{12+\log(\mbox{O}/\mbox{H})}

\newcommand{\ngc}{NGC\,3278}
\newcommand{\sn}{SN\,2009bb}
\urldef{\hyperleda}\url{http://leda.univ-lyon1.fr/ledacat.cgi?NGC%203278}
\newcommand{\grasil}{\sc Grasil}
\newcommand{\magphys}{\sc Magphys}

\begin{document}

 \title{Relativistic supernova 2009bb exploded close to an atomic gas cloud%
 \thanks{Reduced images as FITS files are only available at the CDS via anonymous ftp to \protect\url{cdsarc.u-strasbg.fr (130.79.128.5)} or via
\protect\url{http://cdsarc.u-strasbg.fr/viz-bin/qcat?J/A+A/XXX/AXX}}
 }
 
\titlerunning{SN 2009bb}
\authorrunning{Micha{\l}owski et al.}

\author{Micha{\l}~J.~Micha{\l}owski\inst{\ref{inst:poz},\ref{inst:roe}
}
\and
G.~Gentile\inst{\ref{inst:gent},\ref{inst:brus}}
\and
T. Kr\"uhler\inst{\ref{inst:eso},\ref{inst:mpe}}
\and
H.~Kuncarayakti\inst{\ref{inst:finca},\ref{inst:tuorla}}
\and
P.~Kamphuis\inst{\ref{inst:gmrt}}                                                               
\and
J.~Hjorth\inst{\ref{inst:dark}}                                         
\and
S.~Berta\inst{\ref{inst:zag}}                                           
\and
V.~D'Elia\inst{\ref{inst:elia2},\ref{inst:elia1}}                        
\and
J.~Elliott\inst{\ref{inst:mpe},\ref{inst:hulu}}                                                         
\and
L. Galbany\inst{\ref{inst:pitt}}                                                
\and
J.~Greiner\inst{\ref{inst:mpe}}                                         
\and
L.~K.~Hunt\inst{\ref{inst:hunt}}                                                
\and
M.~P.~Koprowski\inst{\ref{inst:herts}}                          
\and
E.~Le Floc'h\inst{\ref{inst:sacley}}                                    
\and
A.~Nicuesa Guelbenzu\inst{\ref{inst:taut}}                      
\and
E.~Palazzi\inst{\ref{inst:pal}}                                         
\and
J.~Rasmussen\inst{\ref{inst:dark},\ref{inst:dtu}}                       
\and
A.~Rossi\inst{\ref{inst:pal}}                                           
\and 
S.~Savaglio\inst{\ref{inst:sav}}                                                
\and
A.~de Ugarte Postigo\inst{\ref{inst:ant},\ref{inst:dark}}       
\and
P.~van der Werf\inst{\ref{inst:vdw}}                                    
\and
S.~D.~Vergani\inst{\ref{inst:meudon}}                           
        }

\institute{
Astronomical Observatory Institute, Faculty of Physics, Adam Mickiewicz University, ul.~S{\l}oneczna 36, 60-286 Pozna{\'n}, Poland, {\tt mj.michalowski@gmail.com}\label{inst:poz}
\and
SUPA\thanks{Scottish Universities Physics Alliance}, Institute for Astronomy, University of Edinburgh, Royal Observatory, Blakford Hill, Edinburgh, EH9 3HJ, UK
\label{inst:roe}
\and
Sterrenkundig Observatorium, Universiteit Gent, Krijgslaan 281-S9, 9000, Gent, Belgium  \label{inst:gent}
\and
Department of Physics and Astrophysics, Vrije Universiteit Brussel, Pleinlaan 2, 1050 Brussels, Belgium \label{inst:brus}
\and
European Southern Observatory, Alonso de C\'ordova 3107, Vitacura, Santiago, Chile \label{inst:eso}
\and
Max-Planck-Institut f\"{u}r Extraterrestrische Physik, Giessenbachstra{\ss}e, D-85748 Garching bei M\"{u}nchen, Germany \label{inst:mpe}
\and
Finnish Centre for Astronomy with ESO (FINCA), University of Turku, V\"{a}is\"{a}l\"{a}ntie 20, 21500 Piikki\"{o}, Finland \label{inst:finca}
\and
Tuorla Observatory, Department of Physics and Astronomy, University of Turku, V\"{a}is\"{a}l\"{a}ntie 20, 21500 Piikki\"{o}, Finland \label{inst:tuorla}
\and
National Centre for Radio Astrophysics, TIFR, Ganeshkhind, Pune 411007, India \label{inst:gmrt}
\and
Dark Cosmology Centre, Niels Bohr Institute, University of Copenhagen, Juliane Maries Vej 30, DK-2100 Copenhagen \O, Denmark  \label{inst:dark}
\and
Visitor scientist at Department of Physics, Faculty of Science, University of Zagreb, Bijeni\v{c}ka cesta 32, 10000 Zagreb, Croatia \label{inst:zag}
\and
INAF - Osservatorio Astronomico di Roma, Via di Frascati, 33, 00040 Monteporzio Catone, Italy \label{inst:elia2}
\and
ASI Science Data Centre, Via Galileo Galilei, 00044 Frascati (RM), Italy \label{inst:elia1}
\and
Hulu LLC., Beijing, China, 100084 \label{inst:hulu}
\and
PITT PACC, Department of Physics and Astronomy, University of Pittsburgh, Pittsburgh, PA 15260, USA \label{inst:pitt}
\and
INAF-Osservatorio Astrofisico di Arcetri, Largo E. Fermi 5, I-50125 Firenze, Italy \label{inst:hunt}
\and
Centre for Astrophysics Research, University of Hertfordshire, College Lane, Hatfield AL10 9AB, UK \label{inst:herts}
\and
Laboratoire AIM-Paris-Saclay, CEA/DSM/Irfu - CNRS - Universit\'e Paris Diderot, CE-Saclay, pt courrier 131, F-91191 Gif-sur-Yvette, France \label{inst:sacley}
\and
Th\"uringer Landessternwarte Tautenburg, Sternwarte 5, D-07778 Tautenburg, Germany \label{inst:taut}\newpage
\and
INAF-OAS Bologna, Via Gobetti 93/3, I-40129 Bologna, Italy \label{inst:pal}
\and
Technical University of Denmark, Department of Physics, Fysikvej, building 309, DK-2800 Kgs. Lyngby, Denmark \label{inst:dtu}
\and
Physics Department, University of Calabria, I-87036 Arcavacata di Rende, Italy \label{inst:sav}
\and
Instituto de Astrof\' isica de Andaluc\' ia (IAA-CSIC), Glorieta de la Astronom\' ia s/n, E-18008, Granada, Spain \label{inst:ant}
\and
Leiden Observatory, Leiden University, P.O. Box 9513, NL-2300 RA Leiden, The Netherlands \label{inst:vdw}
\and
GEPI, Observatoire de Paris, PSL Research University, CNRS, Place Jules Janssen, 92190 Meudon, France \label{inst:meudon}
}

\abstract{%
}
{The potential similarity of the powering mechanisms of relativistic SNe and GRBs allowed us to make a prediction that relativistic SNe are born in environments similar to those of GRBs, that is, ones which are rich in atomic gas. Here we embark on testing this hypothesis by analysing the properties of the host galaxy NGC 3278 of the relativistic {\sn}.
This is the first time the atomic gas properties of a relativistic SN host are provided and the first time resolved 21\,cm-hydrogen-line ({\hi}) information is provided for a host of an SN of any type in the context of the SN position.
}
{We obtained radio observations with 
ATCA
covering the {\hi} line, and optical integral field unit spectroscopy observations with 
MUSE.
Moreover, we analysed archival carbon monoxide (CO) and multi-wavelength data for this galaxy.
}
{The atomic gas distribution of {\ngc} is not centred on the optical galaxy centre, but instead around a third of atomic gas resides in the region close to the SN position. This galaxy has a few times lower atomic and molecular gas masses than predicted from its star formation rate (SFR). Its specific star formation rate ($\mbox{sSFR}\equiv\mbox{SFR}/\mstar$) is approximately two to three times higher than the main-sequence value.
{\sn} exploded close to the region with the highest SFR density and the lowest age 
($\sim5.5$\,Myr).
Assuming this timescale was the lifetime of the progenitor star, its initial mass would have been close to $\sim36\,\msun$.
}
{As for GRB hosts, the gas properties of {\ngc} are consistent with a recent inflow of gas from the intergalactic medium, which explains the concentration of atomic gas close to the SN position and the enhanced SFR. Super-solar metallicity at the position of the SN (unlike for most GRBs) may mean that relativistic explosions signal a recent inflow of gas (and subsequent star formation), and their type (GRBs or SNe) is determined either 
{\it i)} by the metallicity of the inflowing gas, so that metal-poor gas results in a GRB explosion and metal-rich gas 
results in a relativistic SN explosion without an accompanying GRB, or {\it ii)} by 
the efficiency of gas mixing,
 or {\it iii)} by the type of the galaxy.
 } 


\maketitle

\section{Introduction}
\label{sec:intro}

\begin{figure*}
\begin{center}
\includegraphics[width=\textwidth,clip]{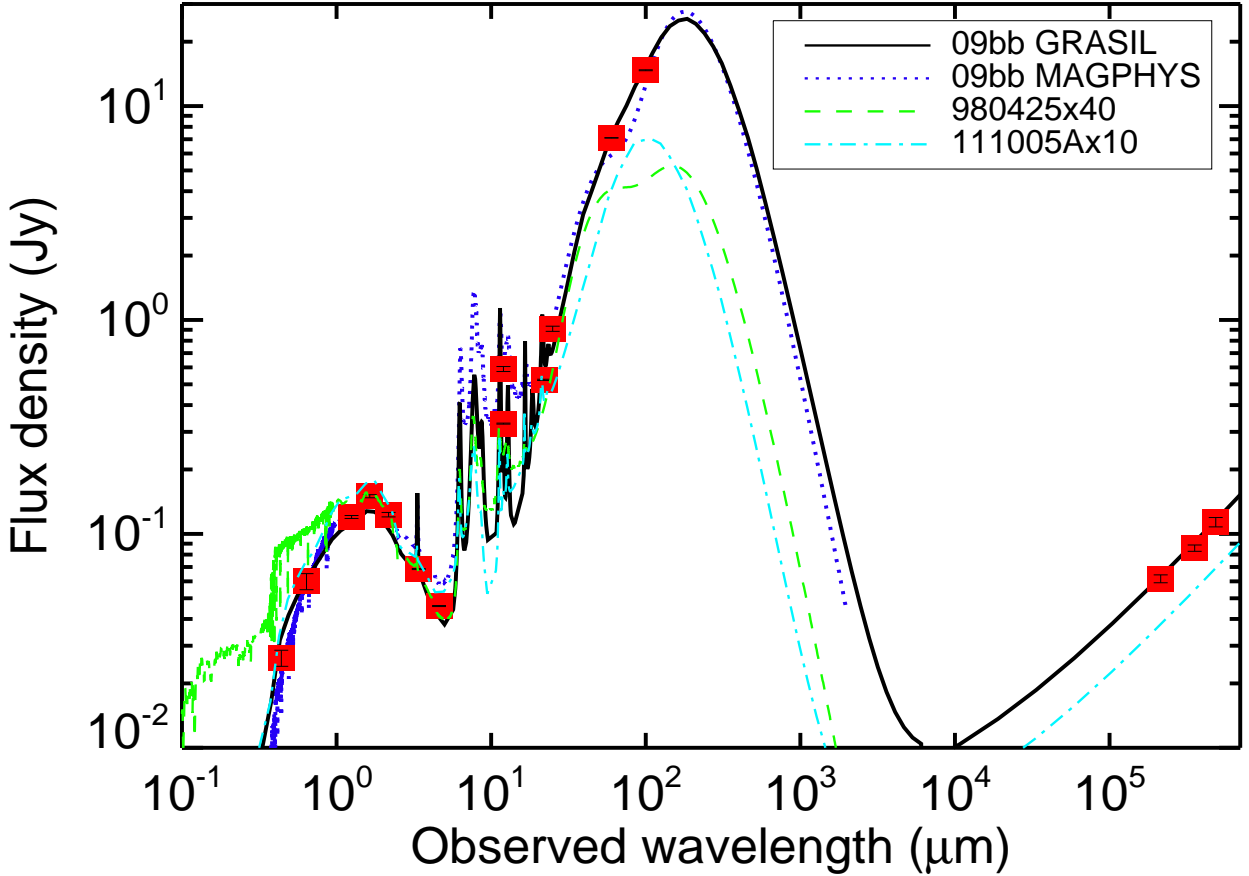}
\end{center}
\caption{Spectral energy distribution of the {\ngc} (red points). The {\sc Grasil} and {\sc Magphys}  models are shown as black  solid and blue dotted lines, respectively. For comparison we show the models for the hosts of GRB 980425 \citep[][]{michalowski14} and 111005A \citep{michalowski18grb}, scaled to approximately match near-IR fluxes.
}
\label{fig:sed}
\end{figure*}

The gas inflow from the intergalactic medium is predicted to be an important process providing the fuel for star formation (see e.g.~\citealt{sancisi08}, \citealt{spring17}). It has been studied mostly from indirect diagnostics because compiling a sample of galaxies for which this process can be observed directly is difficult.
 
Based on the analysis of gas properties in long gamma-ray burst (GRB) host galaxies, we have recently proposed that the progenitors of GRBs are preferentially born when a galaxy accretes fresh gas from the intergalactic medium \citep{michalowski15hi,michalowski16}.  This is based on  a high abundance of atomic gas in GRB hosts and its concentration close to the GRB position \citep{arabsalmani15b,michalowski15hi}.  This may also imply that a fraction of star formation is fuelled directly by atomic, not molecular, gas. The majority of star formation in the Universe is fuelled by molecular gas, as shown by many observations \citep[e.g.][]{carilli13,rafelski16}. However, {\hi}-fuelled star formation has been shown to be theoretically possible  \citep{glover12,krumholz12,hu16} and it was supported by the existence of {\hi}-dominated, star-forming regions in other galaxies \citep{bigiel08,bigiel10,fumagalli08,elmegreen16}. If the connection between GRBs and recent inflow is confirmed, this will allow the use of GRB hosts to study gas accretion and/or {\hi}-fuelled star formation.

On the other hand, relativistic supernovae (SNe) without detected $\gamma$-rays are thought to be powered by similar engines to those of GRBs, but with the jet failing to break out from the exploding star \citep{paragi10,lazzati12,margutti14,chakraborti15,milisavljevic15}. The potential similarity of this powering mechanism to that of GRBs allowed us to make a prediction that relativistic SNe are born in environments similar to those of GRBs, that is, those rich in atomic gas. Here we embark on testing this hypothesis by analysing the properties of the host of the relativistic {\sn}.

\object{SN 2009bb} was discovered by the galaxy-targeted survey, the CHilean Automatic Supernova sEarch (CHASE; \citealt{chase}) on 21 March 2009 \citep{pignata09cbet} at the position of 10:31:33.8762, $-$39:57:30.022 \citep{bietenholz10} and was a broad-line type-Ic supernova \citep{stritzinger09cbet}. 
Radio and optical behaviour, and the relativistic ejecta velocity of {\sn} were very similar to those of low-$z$ GRBs, especially GRB\,980425 
\citep[SN\,1998bw;][]{soderberg10,bietenholz10,pignata11}.

{\sn} exploded within a spiral galaxy type Sa \citep{devaucouleurs91} \object{NGC 3278} (\object{ESO 317-G 043}, \object{PGC 031068}) at a redshift of $0.009877\pm0.000123$ \citep{strauss92}.
It has an inclination to the line of sight of 41\,deg \citep{hyperleda}.\footnote{\hyperleda.}
The {\sn} explosion site was reported to have  super-solar metallicity \citep{levesque10d}.

The objectives of this paper are: {\it i)} to provide the first resolved measurement of the atomic gas properties of a relativistic SN host, {\it ii)} to test whether these properties are consistent with a recent inflow of atomic gas from the intergalactic medium, and {\it iii)} to derive the properties of {\ngc} to assess the possible implications regarding the nature of the progenitor of {\sn}.

We use a cosmological model with $H_0=70$ km s$^{-1}$ Mpc$^{-1}$,  $\Omega_\Lambda=0.7$, and $\Omega_m=0.3$, so {\sn} at $z= 0.009877$ is at a luminosity distance of    42.6 Mpc and $1\arcsec$ corresponds to 203 pc at its redshift. We also assume the 
\citet{chabrier03} 
initial mass function (IMF), to which all star formation rate (SFR) and stellar masses were converted (by dividing by 1.8) if given originally assuming the \citet{salpeter} IMF.

\section{Data}
\label{sec:data}

\begin{figure*}
\begin{center}
\includegraphics[width=\textwidth
]{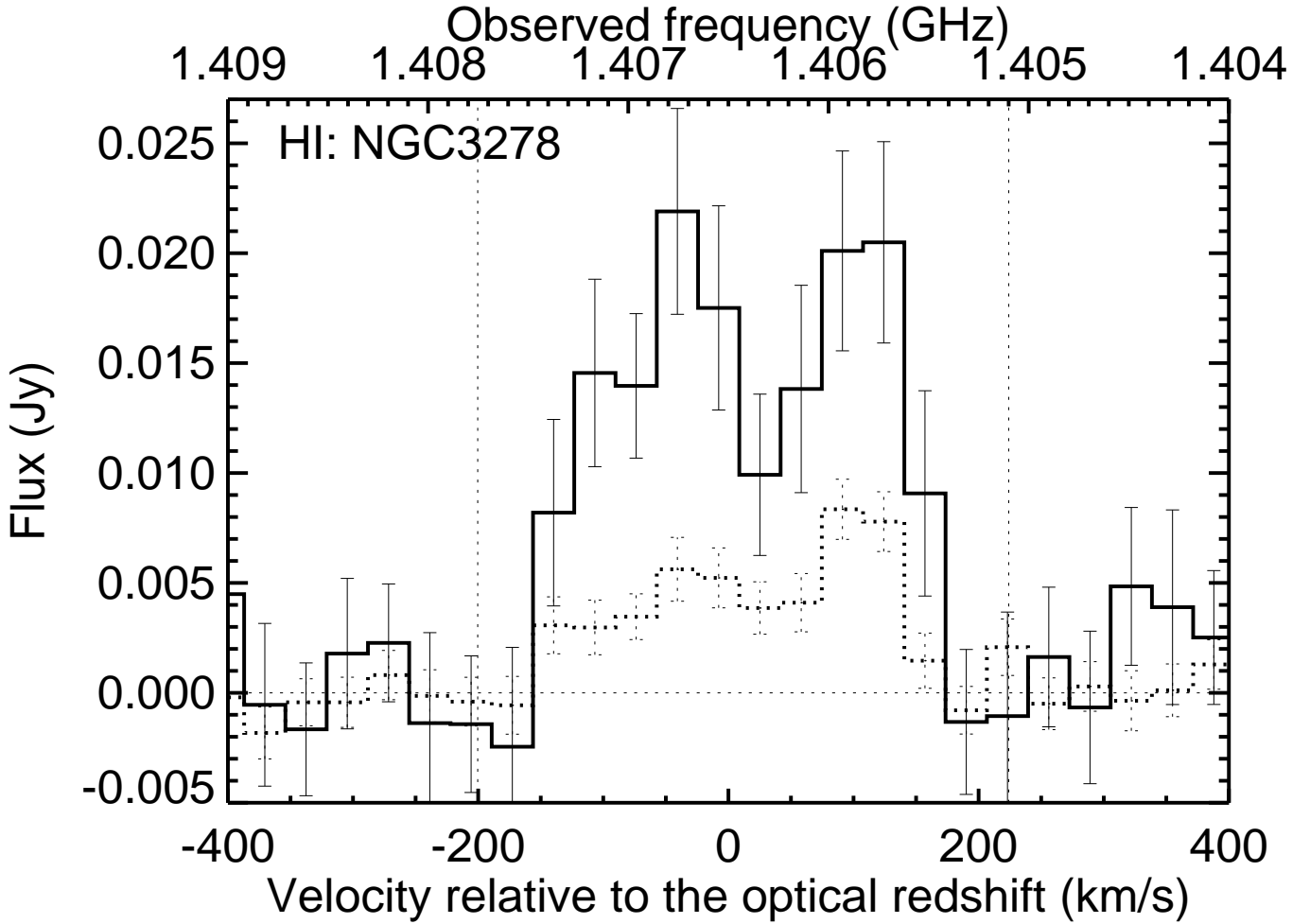} 
\end{center}
\caption{{\hi} spectrum of {\ngc} extracted over the entire galaxy within an aperture of $80\arcsec$ radius (solid histogram) and of the dominant {\hi} region (see Fig.~\ref{fig:himap}) within an aperture of $30\arcsec$ radius   (dotted histogram). 
}
\label{fig:hispec}
\end{figure*}

\begin{figure*}
\begin{center}
\begin{tabular}{cc}
\includegraphics[width=0.5\textwidth]{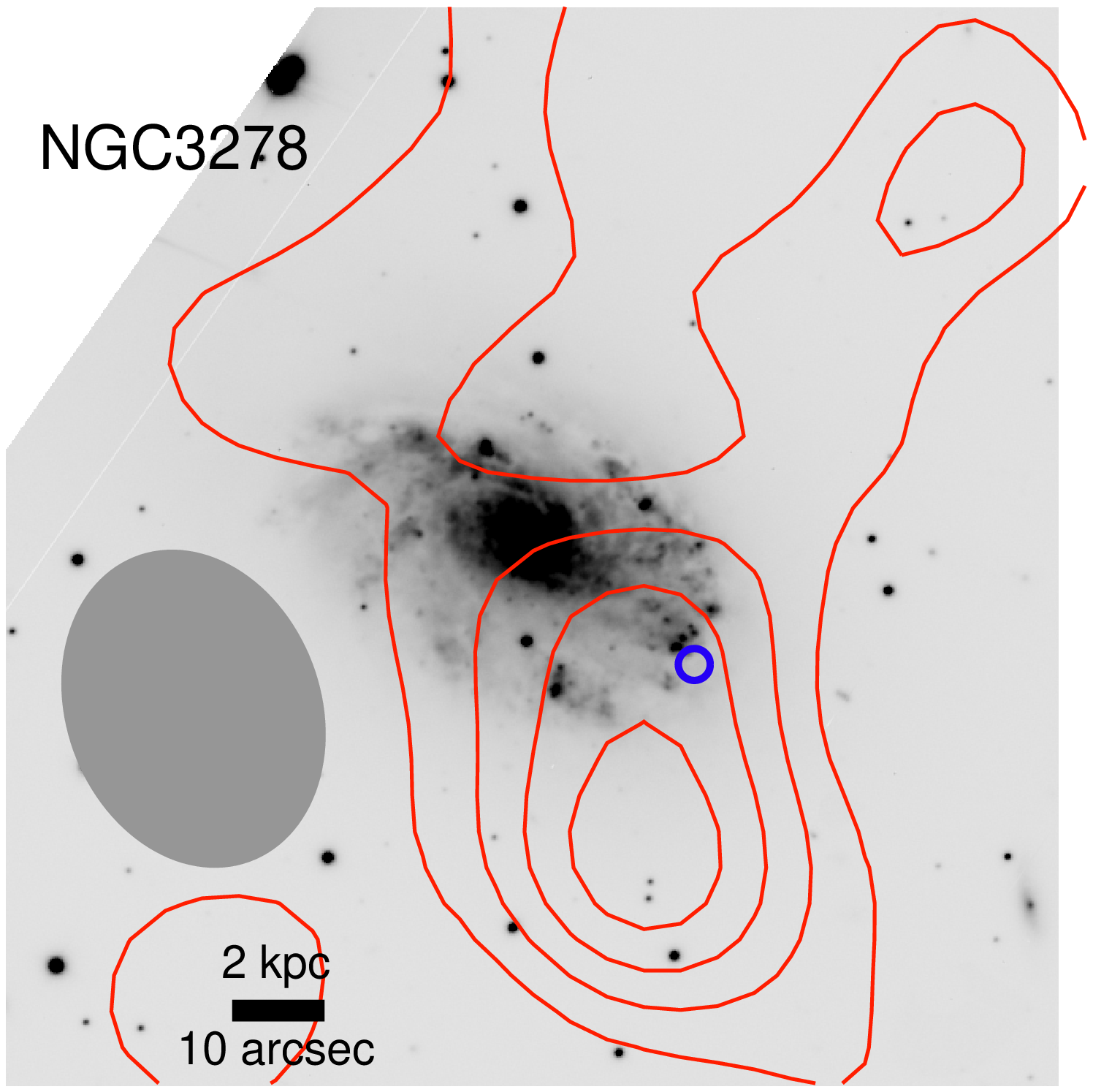} &
\includegraphics[width=0.5\textwidth]{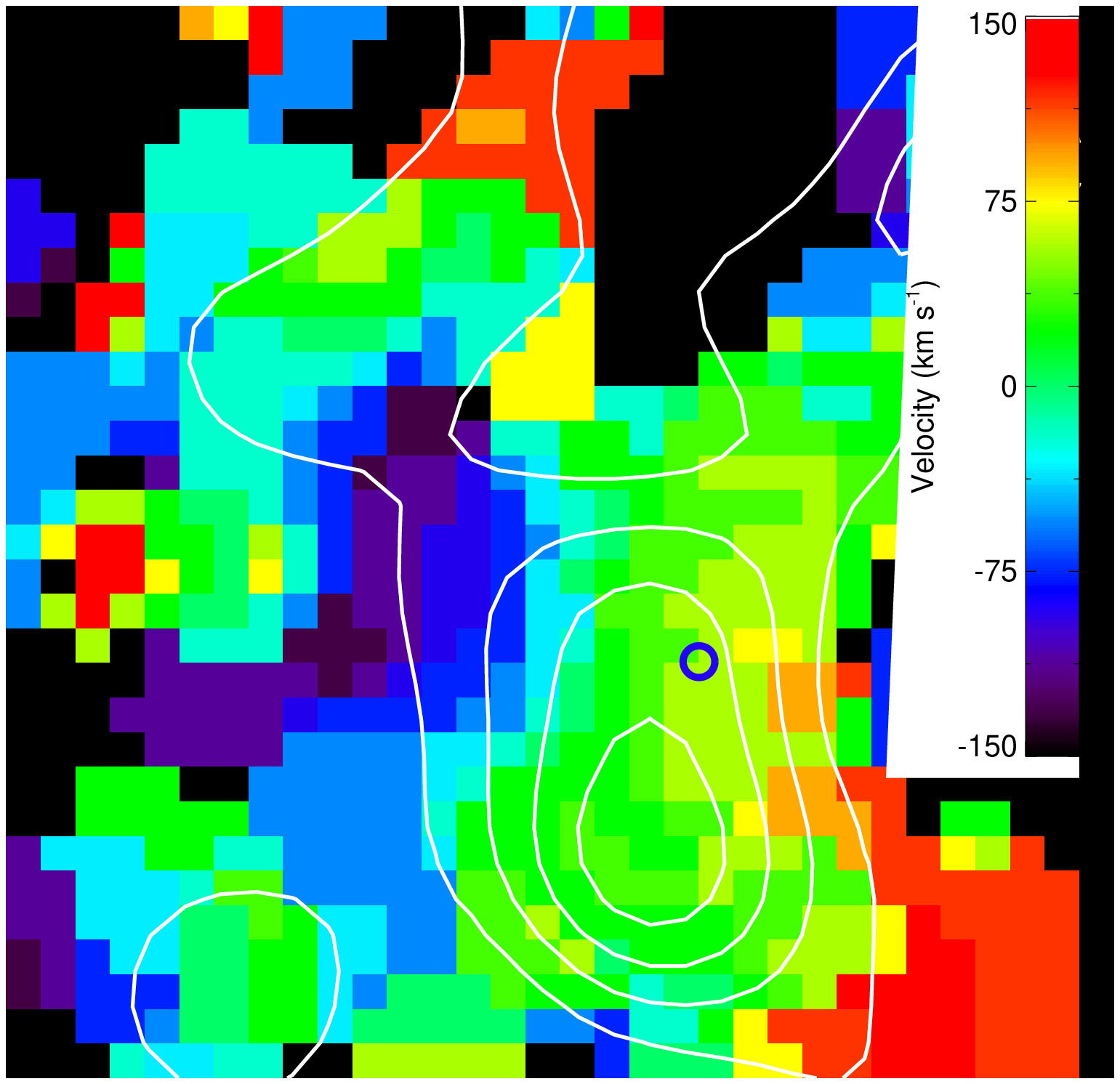} \\
\end{tabular}
\end{center}
\caption{Left: {\hi} contours (red) of {\ngc}  on the optical $r$-band image of the galaxy \citep{pignata11}. The contours are 2, 3, 4, 5$\sigma$, where $\sigma=0.17\,\mbox{Jy beam}^{-1}\,\kms$ (corresponding to a neutral hydrogen column density of $\sim 1.9\times10^{20}\,\mbox{cm}^2$) is the rms of the collapsed image. The position of {\sn} is indicated by the blue circle. {\hi} is concentrated close to this position. The beam size of the {\hi} data is shown as the grey ellipse.  
The image is $120\arcsec\times120\arcsec$ and the scale is indicated by the ruler. North is up and east is left.
Right: The first moment map (velocity field) of the {\hi} line. The image has the same size as the left one and the same contours and the SN position are shown. The velocities are relative to the systemic velocity of $2961\,\kms$ derived from optical spectra.
}
\label{fig:himap}
\end{figure*}

\subsection{Radio}

We performed radio observations with the Australia Telescope Compact Array (ATCA) using the Compact Array Broad-band Backend \citep[CABB;][]{cabb} on 8 March 2016  (project no.~C2700, PI: M.~Micha{\l}owski). 
The array was in the 6B configuration with baselines 214 -- 5939\,m.
The total integration time was $\sim6.5$\,hr.
Sources 1934-638 and 1018-426 were used as the primary and secondary calibrator, respectively.
The data reduction and analysis were done using the {\sc Miriad} package \citep{miriad,miriad2}.

An  intermediate frequency (IF) 
was centred at the {\hi} line in the ATCA CABB `zoom' mode with 32 kHz resolution. We subtracted the continuum to
obtain the continuum-free data, and made the Fourier inversion with the Brigg's weighting robust parameter of 0.5, inverting five channels at a time to get a
data cube with a velocity resolution of 33\,\kms. 
We then made a CLEAN deconvolution down to $\sim$ 3$\sigma$, after which we restored the
channel maps with a Gaussian beam with the size of $36\times28\arcsec$ and a position angle of 19\,deg (from the north towards east). We obtained the rms of $\sim1.5$\,mJy at 33\,{\kms} channels.
We used a $80$\arcsec\  diameter aperture to measure the fluxes for the entire host, and $30$\arcsec\ for the {\hi} peak (see Fig.~\ref{fig:himapwithbox}). 

\subsection{CO}

We used the CO(1-0) and CO(2-1) data obtained with the Swedish European Southern Observatory (ESO) Submillimeter Telescope (SEST) by \citet{albrecht07}. 
The beam sizes are $45$ and $24''$, respectively.
We estimated the molecular gas mass from the CO(1-0) line luminosity 
assuming the Galactic CO-to-{\htwo} conversion factor $\alpha_{\rm CO}=5\msun/(\Kkmspc)$.
For completeness we also estimated the molecular mass from the CO(2-1) line using the flux conversion $S_{\rm CO(1-0)}=0.5\times S_{\rm CO(2-1)}$  \citep[Fig.~4 in][]{carilli13},~$L'_{\rm CO(1-0)}= 2\times L'_{\rm CO(2-1)}$. However, as we possess the CO(1-0) measurement, we do not use the mass based on CO(2-1) in the analysis.

\subsection{Optical integral field spectroscopy}
\label{sec:optifs}

We obtained the observations of {\ngc} using the  Multi Unit Spectroscopic Explorer \citep[MUSE;][]{muse} at the   Very Large Telescope (VLT) on 15 May 2015 (proposal 095.D-0172(A), PI: H. Kuncarayakti, see \citealt{kuncarayakti18} for other results from this programme). The data acquisition and reduction was similar to that described in \citet{kruhler17}. 
The total integration time was $0.5$ hr. The seeing was around $1"$. The data covers a region of $60^{\prime\prime}\times 60^{\prime\prime}$ and the wavelength range $0.475$--$0.93\,\micron$.
To reduce the data, we used the ESO MUSE pipeline\footnote{\urltt{www.eso.org/sci/software/pipelines/.}} 
\citep{musepipe} in the standard manner.
The datacube was corrected for the Milky Way extinction  $E_{B-V} = 0.085$\,mag \citep{schlafly11}.

As in \citet{galbany14}, in order to obtain the galactocentric distance of each pixel, we used the code developed by \citet{krajnovic06}. It analyses the velocity field of the galaxy to obtain the position angle ($\sim156$\,deg for {\ngc}) and the axes ratio ($\sim0.64$). We also derived the inclination to the line of sight of $\sim50\,$deg, close to the value of 41\,deg given by \citet{hyperleda}. In this way the maps of the deprojected distances were obtained and used for radial dependence of estimated properties.

\subsection{Broad-band photometry}

We used the photometry for {\ngc} listed in the NASA/IPAC Extragalactic Database (NED). This includes optical \citep[$B$, $R$;][]{lauberts89}, near-infrared \citep[$J$, $H$, $K$;][]{2mass,2massmain}, mid- and far-infrared  \citep[12, 25, 60, 100\,\micron;][]{sanders03}, and radio \citep[1.4, 0.843\,GHz;][]{condon98,mauch03} data. Additionally we used the 617 MHz flux reported in \citet{soderberg10}

We also used the data from the Wide-field Infrared Survey Explorer (WISE; \citealt{wise}). We used the  fluxes from  the AllWISE Source Catalog\footnote{\url{http://irsa.ipac.caltech.edu/Missions/wise.html}.} 
measured in elliptical apertures with semi-major axes of $31.60$--$33.83\,\arcsec$ ({\tt w[F]gmag}, $\mbox{F}\in 1,2,3,4$), which we list in Table~\ref{tab:wise}.

Finally, we used the VLA 1.4\,GHz continuum data from \citet{condon96}. The image has 18\,{\arcsec} resolution allowing us to investigate the spatial distribution of star formation.

\begin{table}
\caption{WISE fluxes of {\ngc}. \label{tab:wise}}
\begin{tabular}{ccc}
\hline\hline
$\lambda/\micron$ & Flux$/$mJy & aperture/arcsec \\
\hline
3.4      &      $68.47\pm0.38$ & 31.60\\
4.6      &      $45.98\pm0.25$ & 31.60\\
12       &      $327.7\pm1.8$ & 31.60\\
22       &      $524.8\pm3.4$ & 33.83\\
\hline
\end{tabular}
\end{table}

\renewcommand{\tabcolsep}{0.1cm}
\begin{table*}
\caption{{\sc Magphys} results from the SED fitting. \label{tab:magphysres}}
\scriptsize
\begin{center}
\begin{tabular}{ccccccccccccccc}
\hline\hline
$\log L_{\rm IR}$ & SFR & sSFR & $\log M_*$ & $\log M_d$ & $\tau_V$ & $T_{\rm cold}$ & $\xi_{\rm cold}$ & $T_{\rm warm}$ & $\xi_{\rm warm}$ & $\xi_{\rm hot}$ & $\xi_{\rm PAH}$ &  $f_\mu$ & $\log\mbox{age}_M$\\
($L_\odot$) & ($M_\odot\,\mbox{yr}^{-1}$) & (Gyr$^{-1}$) & ($M_\odot$) & ($M_\odot$) & & (K) & & (K) & & & & & (yr)\\
(1) & (2) & (3) & (4) & (5) & (6) & (7) & (8) & (9) & (10) & (11) & (12) & (13) & (14)\\
\hline
 $10.72^{+0.06}_{-0.03}$ & $2.77^{+0.88}_{-0.85}$ & $0.08^{+0.07}_{-0.03}$ & $10.54^{+0.12}_{-0.15}$ & $7.97^{+0.48}_{-0.44}$ & $2.42^{+1.23}_{-0.84}$ & $18.9^{+3.2}_{-2.5}$ & $0.43^{+0.04}_{-0.03}$ & $55^{+4}_{-3}$ & $0.30^{+0.04}_{-0.03}$ & $0.11^{+0.02}_{-0.02}$ & $0.154^{+0.021}_{-0.022}$ & $0.66^{+0.08}_{-0.08}$ & $9.84^{+0.10}_{-0.10}$ &  \\
\hline
\end{tabular}
\tablefoot{(1) $8-1000\,\mu$m infrared luminosity. (2) Star formation rate from SED modelling. (3) Specific star formation rate ($\equiv\mbox{SFR}/M_*$). (4) Stellar mass. (5) Dust mass. (6) Average $V$-band optical depth ($A_V=1.086\tau_V$).  (7) Temperature of the cold dust component. (8) Contribution of the cold component to the infrared luminosity. (9) Temperature of the warm dust component. (10) Contribution of the warm component to the infrared luminosity. (11) Contribution of the hot ($130$--$250$ K, mid-IR continuum) component to the infrared luminosity. (12) Contribution of the PAH component to the infrared luminosity. (13) Contribution of the ISM dust (as opposed to birth clouds) to the infrared luminosity. (14) Mass-weighted age.}
\end{center}
\end{table*}

\renewcommand{\tabcolsep}{0.1cm}
\begin{table*}
\caption{{\sc Grasil} results from the SED fitting. \label{tab:grasilres}}
\scriptsize
\begin{center}
\begin{tabular}{cccccccccc}
\hline\hline
 $\log L_{\rm IR}$ & $\mbox{SFR}_{\rm IR}$ & $\mbox{SFR}_{\rm SED}$ & $\mbox{SFR}_{\rm UV}$ & $\mbox{sSFR}_{\rm SED}$ & $\log M_*$ & $\log M_{\rm dust}$ & $\log T_{\rm dust}$ & $A_V$ & $\log\mbox{age}_M$ \\
 ($L_\odot$) & ($M_\odot\,\mbox{yr}^{-1}$) & ($M_\odot\,\mbox{yr}^{-1}$) & ($M_\odot\,\mbox{yr}^{-1}$) & ($\mbox{Gyr}^{-1}$) & ($M_\odot$) & ($M_\odot$) & (K) & (mag) & (yr) \\
 (1) & (2) & (3) & (4) & (5) & (6) & (7) & (8) & (9) & (10) \\
\hline
 10.74 & 5.21 & 3.69 & 0.54 & 0.14 & 10.43 & 8.93 & 21 & 0.96 & 9.93 \\
\hline
\end{tabular}
\tablefoot{ (1) $8$--$1000\,\mu$m infrared luminosity. (2) Star formation rate from $L_{\rm IR}$ \citep{kennicutt}. (3) Star formation rate from SED modelling. (4) Star formation rate from UV emission \citep{kennicutt}. (5) Specific star formation rate ($\equiv\mbox{SFR}_{\rm SED}/M_*$). (6) Stellar mass. (7) Dust mass. (8) Dust temperature. (9) Mean dust attenuation at $V$-band. (10) Mass-weighted age. 
}
\end{center}
\end{table*}

\begin{table*}
\small
\centering
\caption{{\hi} properties of NGC\,3278.\label{tab:mhi}}
\medskip
\begin{tabular}{lccccccc}
\hline
Region & \zhi & $v_{\rm HI}$ & $v_{\rm FWHM}$ & $F_{\rm peak}$ & $F_{\rm int}$   &  $\log(\lphi)$          & $\log(\mhi)$ \\
    &      & (km s$^{-1}$) & (km s$^{-1}$)  & (mJy)           & (Jy km s$^{-1}$) & ($\mbox{K km s}^{-1} \mbox{ pc}^2$) & (${\rm M}_\odot$) \\
(1)    & (2)     & (3)  & (4)           & (5) & (6) & (7) & (8) \\
\hline
NGC\,3278 & $0.009890 \pm        0.000040$ & $2965\pm      12$ &       $222 \pm 27$ & $19.2 \pm 1.9$ & $4.84 \pm 0.53$ & $11.148 \pm 0.045$ & $9.318 \pm 0.045$   \\
{\hi} peak & $0.009983 \pm        0.000044$ & $2993 \pm     13$& $243 \pm 16$ & $5.9 \pm 0.6$ & $1.53 \pm 0.17$ & $10.657 \pm 0.045$ & $8.827 \pm 0.045$   \\
\hline
\end{tabular}
\tablefoot{
(1) Either the entire galaxy (aperture radius of $80\,\arcsec$) or the {\hi} peak (see Fig.~\ref{fig:himap}, aperture radius of $30\,\arcsec$). (2) Redshift determined from the Gaussian fit to the {\hi} spectrum. (3) The corresponding velocity (4) Full width at half maximum of this Gaussian. (5) Peak of this Gaussian. (6) Integrated flux within $2\sigma$ of the Gaussian width. (7) {\hi} line luminosity using Eq. 3 in \citet{solomon97}. (8) Neutral hydrogen mass  using Eq. 2 in \citet{devereux90}.
}
\end{table*}

\begin{table*}
\centering
\caption{CO fluxes, luminosities, and molecular gas masses of NGC 3278, based on the data from \citet{albrecht07}.\label{tab:co}}
\medskip
\begin{tabular}{lccccc}
\hline
CO & FWHM &  $F_{\rm int}$   &  $\log(L'_{\rm CO})$           & $\log(L_{\rm CO})$ & $\log(\mhtwo)$  \\
trans. & (\arcsec) & (Jy km s$^{-1}$) & ($\mbox{K km s}^{-1} \mbox{ pc}^2$) & ($\lsun$) & ($\msun$)  \\
(1)    & (2)     & (3)  & (4) & (5)  & (6)         \\
\hline
1-0 & 45 & 132 & 8.75 & 4.41 & 9.45 \\ 
2-1 & 24 & 184 & 8.29 & 4.86 & 9.29 \\ 
\hline
\end{tabular}
\tablefoot{
(1) CO transition. (2) Full width at half maximum of the telescope (green circles on Fig.~\ref{fig:himapwithbox}). (3) Integrated flux. (4) Line luminosity using Eq. 3 in \citet{solomon97}. (5) Line luminosity in solar luminosity units. (6) Molecular hydrogen mass assuming CO-to-{\htwo} conversion factor $\alpha_{\rm CO}=5\msun/(\Kkmspc)$. In order to calculate {\mhtwo} from the 2-1 transition, we assumed $L'_{\rm CO(1-0)}= 2\times L'_{\rm CO(2-1)}$.
}
\end{table*}

\begin{figure}
\includegraphics[width=0.5\textwidth]{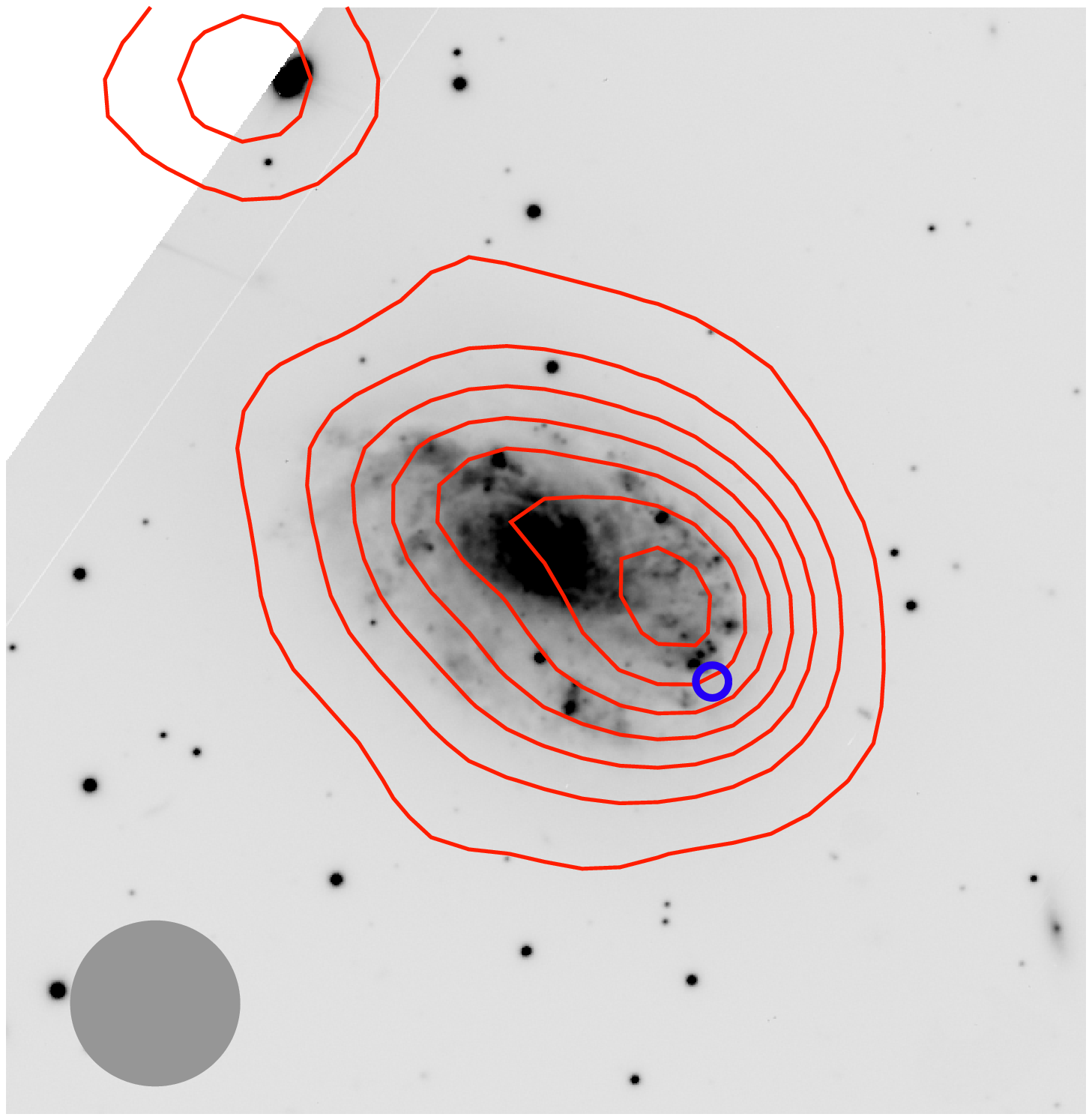}
\caption{Continuum 1.4\,GHz contours (red; from \citealt{condon96}) of {\ngc}  on the optical $r$-band image of the galaxy \citep{pignata11}. The lowest contour is at 1\,mJy\,beam$^{-1}$ and the step is 2\,mJy\,beam$^{-1}$.
The position of {\sn} is indicated by the blue circle. Radio continuum emission peaks close to this position. The beam size of the radio data is shown as the grey circle.  
The image is $120\arcsec\times120\arcsec$. 
North is up and east is left.
}
\label{fig:contmap}
\end{figure}

\section{SED modelling}
\label{sec:method}

For the host galaxy emission, we applied the spectral energy distribution (SED) fitting method detailed in \citet[][see therein a discussion of the derivation of galaxy properties and typical uncertainties]{michalowski08,michalowski09,michalowski10smg,michalowski10smg4,michalowski12mass,michalowski14mass}, which is based on 35\,000 templates from the library of \citet{iglesias07} plus some templates from \citet{silva98} and \citet{michalowski08}, all of which were developed using {\grasil}\footnote{\url{adlibitum.oats.inaf.it/silva/grasil/grasil.html}.} \citep{silva98}. They are based on numerical calculations of radiative transfer within a galaxy, which is assumed to be a triaxial axisymmetric system with diffuse dust and dense molecular clouds in which stars are born.

The templates cover a broad range of galaxy properties from quiescent to starburst, and span an $A_V$ range from $0$ to $5.5$ mag. The extinction curve \citep[Fig.~3 of][]{silva98} is derived from the modified dust grain size distribution of \citet{draine84}.
The star formation histories are assumed to be a smooth Schmidt-type law \citep[i.e. the SFR is proportional to the gas mass;
see][for details]{silva98} with a starburst (if any) on top of that, starting $50$ Myr before the time at which the SED is computed. There are seven free parameters in the library of \citet{iglesias07}: the normalisation of the Schmidt-type law, the timescale of the mass infall, the intensity of the starburst, the timescale for molecular cloud destruction, the optical depth of the molecular clouds, the age of the galaxy, and the inclination of the disk with respect to the observer.

We also used {\magphys}\footnote{\url{www.iap.fr/magphys}.} \citep[Multi-wavelength Analysis of Galaxy Physical Properties;][]{dacunha08}, which is an empirical, physically-motived SED modelling code that is based on the energy balance between the energy absorbed by dust and that re-emitted in the infrared. We used the \citet{bruzualcharlot03} stellar population models and adopted the \citet{chabrier03} IMF. 

Similarly to {\grasil}, in {\magphys} two dust media are assumed: a diffuse interstellar medium (ISM) and dense stellar birth clouds. Four dust components are taken into account: cold dust ($15$--$25$ K), warm dust ($30$-$60$ K), hot dust ($130$--$250$ K), and polycyclic aromatic hydrocarbons (PAHs). A simple power-law attenuation law is assumed.

\section{Results}
\label{sec:res}

\subsection{Integrated stellar properties}
\label{sec:stell}

The best-fit SED models are presented in Fig.~\ref{fig:sed} compared with the host galaxies of GRB\,980425 \citep{michalowski14} and 111005A \citep{michalowski18grb}. The derived galaxy properties are listed in Tables~\ref{tab:magphysres} and \ref{tab:grasilres}. Our derived SFR is consistent with that reported in \citet{levesque10d}. 

In terms of stellar mass, {\ngc} is a typical galaxy with  $\mstar\sim3\times10^{10}\,\msun$, close to the knee of the stellar mass function of  local spiral galaxies \citep{moffett16}. 
However, the specific SFR ($\mbox{sSFR}\equiv\mbox{SFR}/\mstar$) of $0.08$--$0.14\,\mbox{Gyr}^{-1}$ is approximately two to three times higher than the main-sequence value of $0.04$--$0.05\,\mbox{Gyr}^{-1}$ at this redshift and mass  \citep{speagle14}. Hence, {\ngc} is at the higher end of the main sequence towards starburst galaxies. The stellar mass of {\ngc} is at least an order of magnitude higher than those of low-$z$ GRB hosts (\citealt{savaglio09,castroceron10,vergani15,japelj16,perley16b}, but see an atypical massive and quiescent host presented by \citealt{rossi14}).
 
 Both models return quite high levels of visual dust attenuation ($A_V\sim1$--$2$\,mag), which is evidenced by red optical colours, similar to those of the GRB\,111005A host; however, that galaxy is nearly edge-on, whereas {\ngc} has an inclination of 41\,deg, so the amount of dust is much higher. Indeed the dust mass of {\ngc} is approximately two orders of magnitude higher than that of the GRB\,111005A host (whereas the stellar mass of the latter is less than a factor of ten smaller). 
 The dust mass estimates of {\ngc} with {\sc Grasil} and {\sc Magphys}  differ by an order of magnitude, but the lack of long-wavelength data above $100\,\micron$ means that this parameter is very poorly constrained.
 A factor of two difference is due to different mass absorption coefficients $\kappa$ and the rest is due to differences in assumed temperatures and distributions of dust components.

\subsection{Gas properties}
\label{sec:gas}

The {\hi} fluxes at each frequency element were determined by aperture photometry with the aperture radius of $80''$ for the entire galaxy and of $30''$ for the {\hi} peak. 
When we, instead, fit a two-dimensional (2-D) Gaussian with the size of the beam at the position of the {\hi} peak, we obtained the {\hi} flux of $\sim1.30\pm0.17$\,Jy\,\kms, consistent with the aperture estimate of $\sim1.53\pm0.17$\,Jy\,\kms.

The spectra are shown in Fig.~\ref{fig:hispec}. Gaussian functions were fitted to them and the parameters of the fit are reported in columns 2--4 of Table~\ref{tab:mhi}.  The {\hi} emission map derived from the collapsed cube within $2\sigma$ from this fit 
 (dotted lines in Fig.~\ref{fig:hispec})
is shown in  Fig.~\ref{fig:himap}. This range was also used to obtain integrated {\hi} emission ($F_{\rm int}$ in Jy\,\kms) directly from the spectra (not from the Gaussian fit, which is not a perfect representation of the line shape). The line luminosity ({\lphi} in K\,\kms pc$^2$) was calculated using Eq. 3 in \citet{solomon97} and transformed to {\mhi} using Eq. 2 in \citet{devereux90}. 
For the HI peak the flux corresponds to the neutral hydrogen column density of $\sim(1.67\pm  0.19)\times10^{21}\,\mbox{cm}^2$.

The {\hi} first moment map (velocity field) is shown on the right panel of Fig.~\ref{fig:himap}. The large beam does not allow detailed velocity analysis, but the field does not resemble clearly a rotating disk (positive velocities are on both sides of regions with negative velocities).
On the other hand, the {\hi} spectrum exhibits a double-peaked profile characteristic of a rotating disk, but the significance of this feature is low. Therefore it is likely that only a fraction of atomic gas is within a rotating disk giving rise to this  double-peaked profile.

We detected and resolved the {\hi} emission of the target, so that we are able to identify the main concentration of atomic gas.
This is the first time {\hi} data for a relativistic SN host is provided and the first time resolved {\hi} information is analysed for the host of an SN of any type in the context of the SN position \citep[non-resolved results were presented in][]{galbany17}.

The emission is not concentrated near the galaxy centre, but towards the SN position (the peak is $\sim21"$ [$\sim4$\,kpc], i.e.~one beam, south of the SN position). This concentration is responsible for $\sim32 \pm 5$\% of the total integrated flux. 
The remaining emission probably comes from a rotating disk (giving rise to the double-peaked {\hi} profile in Fig.~\ref{fig:hispec}). It seems that the sensitivity of our data allowed us to clearly detect only the strongest concentration of {\hi}, leaving the emission from the disk difficult to identify.

Given limited $uv$-coverage, we investigated the issue of whether we resolve out a significant fraction of the {\hi} flux. Our observations in the ATCA 6B configuration are limited by a largest recoverable scale of $\sim105"$
%
(Table 1.5 in the ATCA Users Guide%
\footnote{\urltt{www.narrabri.atnf.csiro.au/observing/users\_guide/ html/atug.html.}}, a more optimistic estimate based on the ratio of the observed wavelength and the shortest baseline of 214\,m gives $\sim200"$).
This is larger than the optical extent of the galaxy (diameter of $\sim60"\times40"$), so our observations are unlikely to resolve out a lot of {\hi} emission. Even if the atomic gas disk is a few times larger than the optical disk (which is not uncommon) and we do resolve out some of the extended emission, then our conclusion is still valid that the strongest atomic gas concentration is located away from the galaxy centre towards the SN position. A similar strong {\hi} concentration away from the galaxy centre was detected for the GRB\,980425 host by the Giant Metrewave Radio Telescope \citep[GMRT;][]{arabsalmani15b}.

However, it is unlikely that we resolve out a significant fraction of the total emission, because our measurement agrees with the single-dish flux.
\citet{courtois11} and \citet{roth94} provided low resolution Green Bank Telescope and Parkes {\hi} data for {\ngc} (as a part of a larger survey of local galaxies, so they did not discuss that this was a relativistic SN host). They reported the linewidth at a flux level that is 50\% of the mean flux averaged in channels within the wavelength range enclosing 90\% of the total integrated flux, $W_{m50}=292\pm11$ and $295\,\kms$, respectively. This is slightly higher than the Full width at half maximum (FWHM) given in Table~\ref{tab:mhi} because the  Gaussian function does not represent the line profile accurately. Indeed the Gaussian FHWM reported by \citet{roth94} of $225\,\kms$ is consistent with our result. Our estimate of the integrated flux (which does not involve assumptions on the line shape), agrees with $5.9\pm0.5$\,Jy\,\kms and $6.9\pm3.3$\,Jy\,{\kms} reported by \citet{courtois11} and \citet{roth94}, respectively.

According to the SFR-{\mhi} scaling relation \citep[Eq.~1 in][]{michalowski15hi}, {\ngc} with $\mbox{SFR}\sim3\,\msunyr$ (Tables~\ref{tab:magphysres} and \ref{tab:grasilres}) should have $\log(\mhi/\msun)\sim10$,~$0.7$ dex higher than the measured value. The scatter of this relation is significant ($0.38$\,dex $1\sigma$), so this is not unusual, but we conclude that {\ngc} exhibits low atomic gas content for its SFR.

The CO fluxes, luminosities, and the resulting molecular gas masses are presented in Table~\ref{tab:co}. 
The molecular gas mass based on the CO(2-1) line is a $\sim0.15\,$dex lower than that based on the CO(1-0), but this is very likely due to the beam size at the CO(2-1) transition of 24{\arcsec} being too small to cover the entire galaxy (Fig.~\ref{fig:himapwithbox}), so the corresponding CO(2-1) flux is underestimated. 
Therefore, only the estimates based on the CO(1-0) line are used in the following analysis.

Using the total infrared luminosity of $L_{\rm IR}\sim5\times10^{10}\,\lsun$ (Tables~\ref{tab:magphysres} and \ref{tab:grasilres}), we estimate the star formation efficiency  (SFE) of $L_{\rm IR}/L'_{\rm CO(1-0)}\sim100\,\lsun/(\Kkmspc)$.
This is one of the highest numbers among local spirals with $\sim(48\pm7)\,\lsun/(\Kkmspc)$ derived by \citet[][their Fig.~13]{daddi10}. 
Similarly, the relation between SFR, CO luminosity, and metallicity presented in \citet[][their Fig.~5]{hunt15}, $\log(\mbox{SFR}/\lp_{\rm CO}) =-2.25\times[12+\log(\mbox{O}/\mbox{H})]+11.31$ predicts an $\mbox{SFR}/\lp_{\rm CO}$ of the {\sn} host of $\sim3.5\times10^{-9}\,\msunyr / (\Kkmspc)$, whereas the measured value is $\sim1.5$--$2$ times higher, $5$--$6.5\times10^{-9}\,\msunyr / (\Kkmspc)$, indicating low CO luminosity for its SFR and metallicity.
Hence, the {\sn} host galaxy has also a few times lower molecular gas content than its SFR would suggest. Molecular gas deficiency was also claimed for some GRB hosts (\citealt{hatsukade14,stanway15,michalowski16,michalowski18co}, but we note that this result does not hold for the host galaxy of GRB\,020819B, for which the initial host identification was proven to be wrong; see \citealt{perley17}). 
On the other hand, normal molecular gas properties were found in other GRB hosts \citep{arabsalmani18}, with the current status that the sample on average does not deviate from other star-forming galaxies \citep{michalowski18co}.

We also used the relation between the metallicity, atomic gas, and molecular gas for dwarf galaxies  provided by \citet[][their Sect. 4]{filho16}, based on the calibration of \citet{amorin16}: $\log({\mhtwo}) = 1.2 \log({\mhi}) - 1.5\times [12 + \log(\mbox{O}/\mbox{H}) - 8.7] - 2.2$. For its atomic gas mass (Table~\ref{tab:mhi}) and average metallicity (Table~\ref{tab:values}, last row using the calibration of \citealt{dopita16}), the {\sn} host should have $\log(\mhtwo/\msun)\sim8.85$, approximately four times lower than the actual CO estimate (Table~\ref{tab:co}). 
{\ngc} would be consistent with this relation if it had a much lower metallicity of $\metoh\sim8.4$ (half solar where solar metallicity is $\metoh\sim8.66$ \citealt{asplund04}).

We find that the {\sn} host  has a molecular gas mass fraction of $\mhtwo/(\mhtwo + \mhi)\sim57$\%, which is high, but within the range for other star-forming galaxies \citep[a percentage of a few to a few tens;][]{young89b,devereux90,leroy08,saintonge11,cortese14,boselli14}, and of hosts of SNe of different type \citep{galbany17}.

\subsection{Resolved ISM and stellar properties}

\begin{table}
\caption{Linear fit of the properties as a function of distance from the galaxy centre (Fig.~\ref{fig:musedist}) in the form $A+B_1 \times\mbox{dist}_{\rm arcsec}$ or $A+B_2 \times\mbox{dist}_{\rm kpc}$. The variables and the units are as in Table~\ref{tab:values}.  \label{tab:distfit}}
\begin{tabular}{lccc}
\hline
\hline
Property & $A$ & $B_1$ & $B_2$ \\
         & (unit) & (unit/arcsec) & (unit/kpc) \\
\hline
\verb!HaFlux! & 11.8 & 0.72 & 3.56 \\
\verb!HaEW! & 7.6 & 4.8 & 23.6 \\
\verb!SFR! & 0.27 & 0.0165 & 0.081 \\
\verb!Mass! & 0.05 & -0.0023 & -0.011 \\
\verb!sSFR! & -0.00112 & 0.0022 & 0.011 \\
\verb!ebv! & 0.84 & -0.017 & -0.086 \\
\verb!OH_D16! & 8.97 & -0.015 & -0.072 \\
\verb!OH_PP04_03N2! & 8.83 & -0.006 & -0.028 \\
\verb!OH_PP04_N2! & 8.73 & -0.005 & -0.025 \\
\hline
\end{tabular}
\end{table}

\begin{figure*}
\begin{center}
\begin{tabular}{cc}
\includegraphics[width=0.45\textwidth]{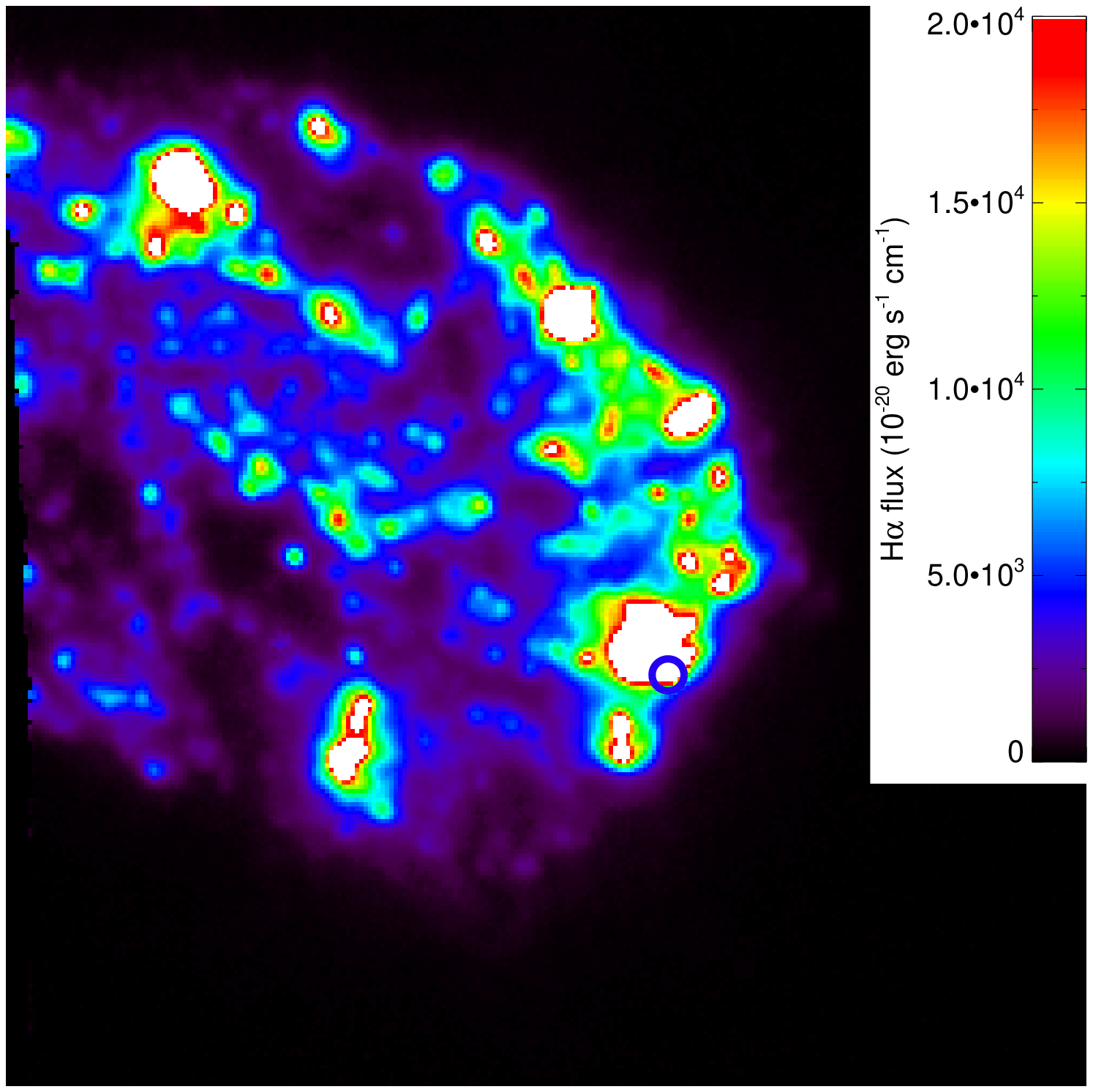} & 
\includegraphics[width=0.45\textwidth]{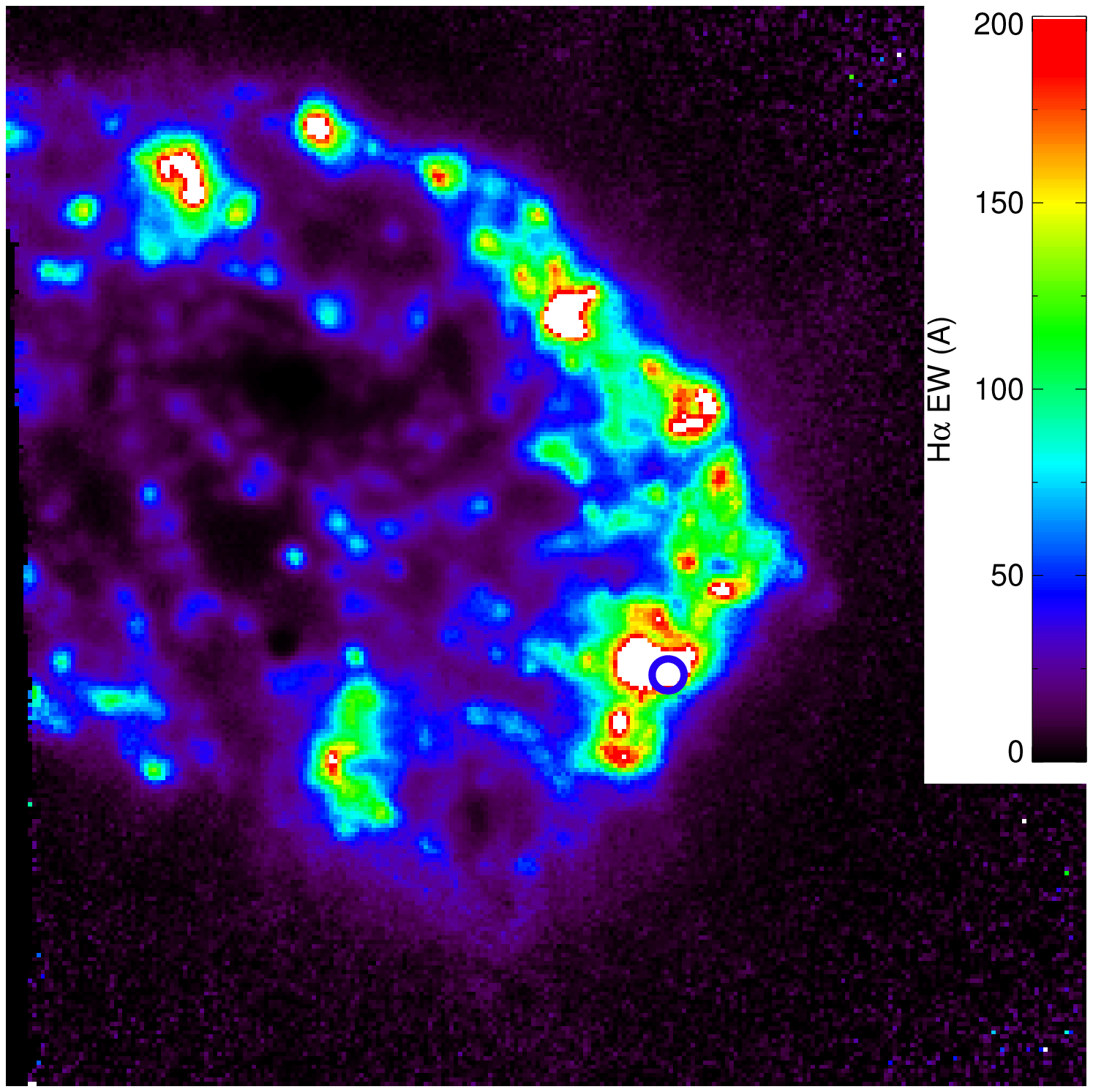} \\
\includegraphics[width=0.45\textwidth]{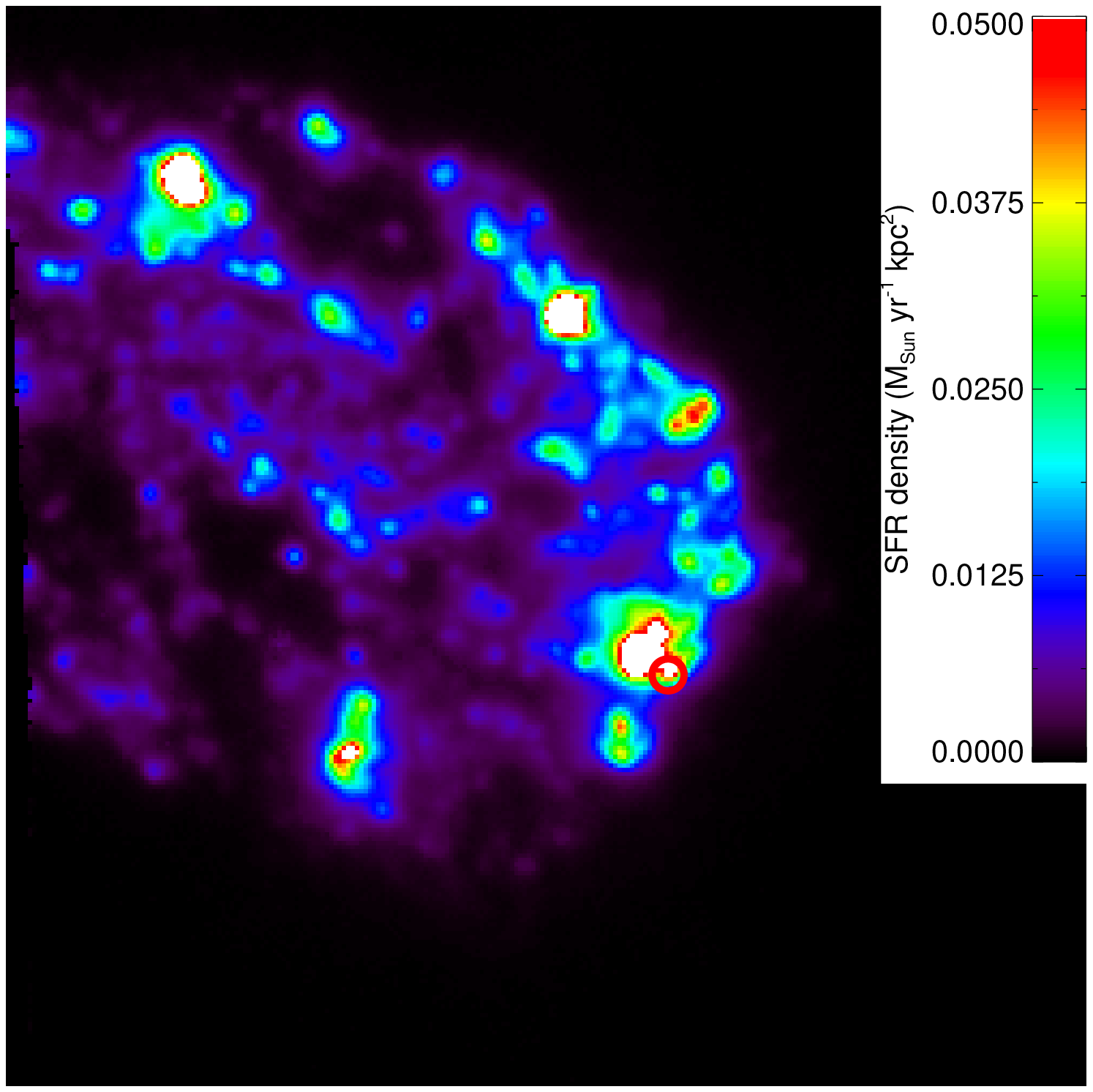} & 
\includegraphics[width=0.45\textwidth]{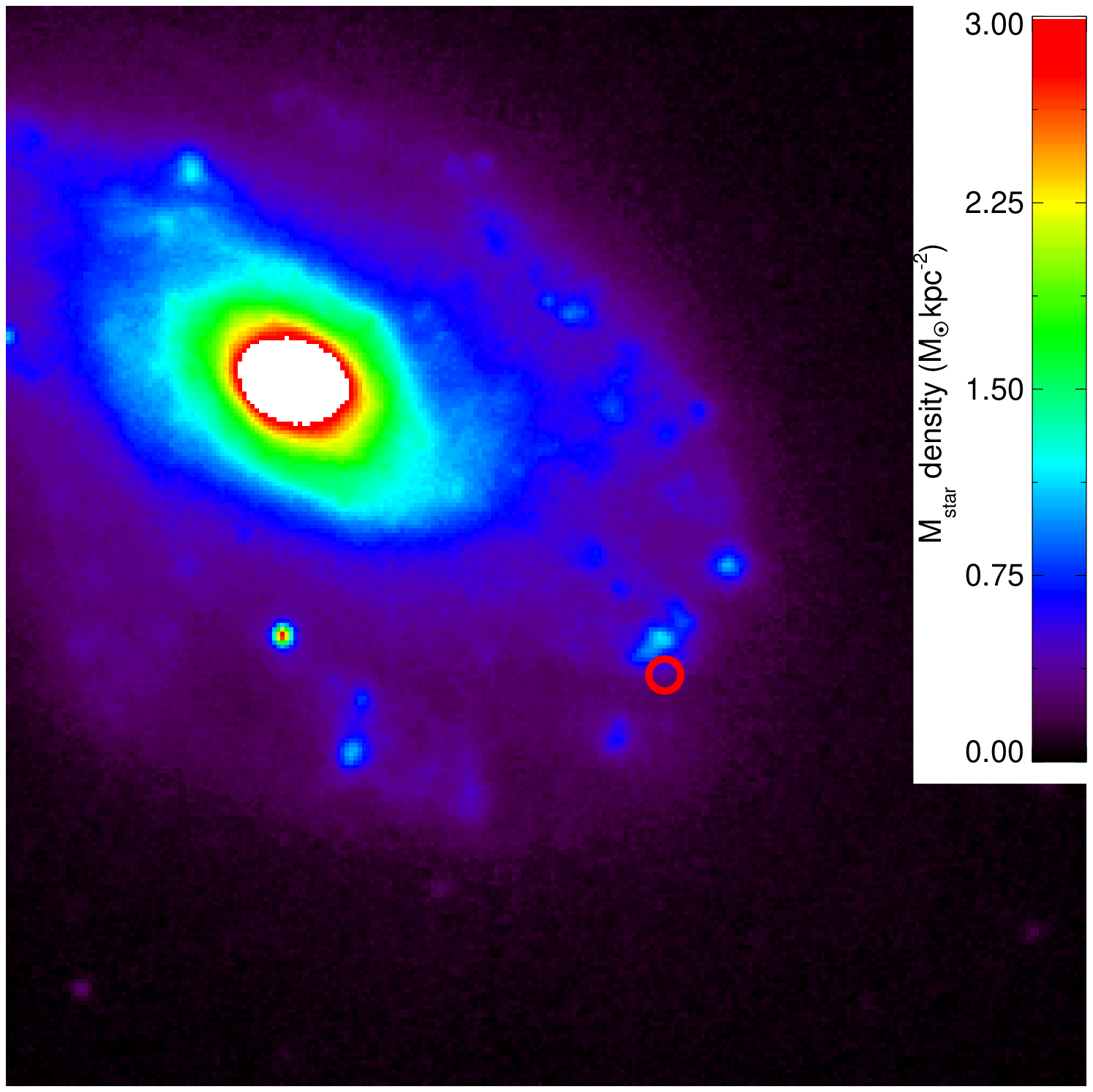} \\
\includegraphics[width=0.45\textwidth]{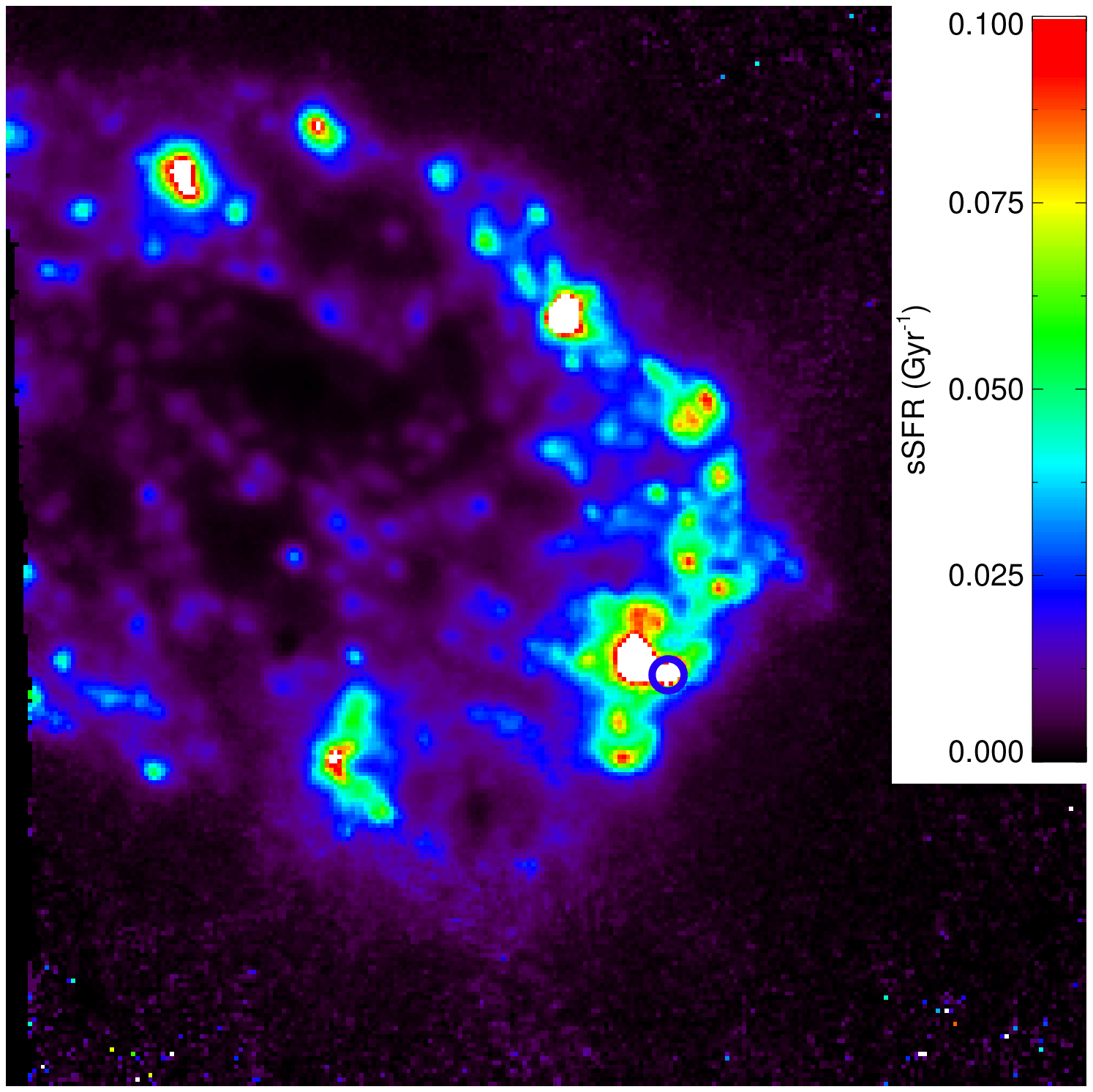} &
\includegraphics[width=0.45\textwidth]{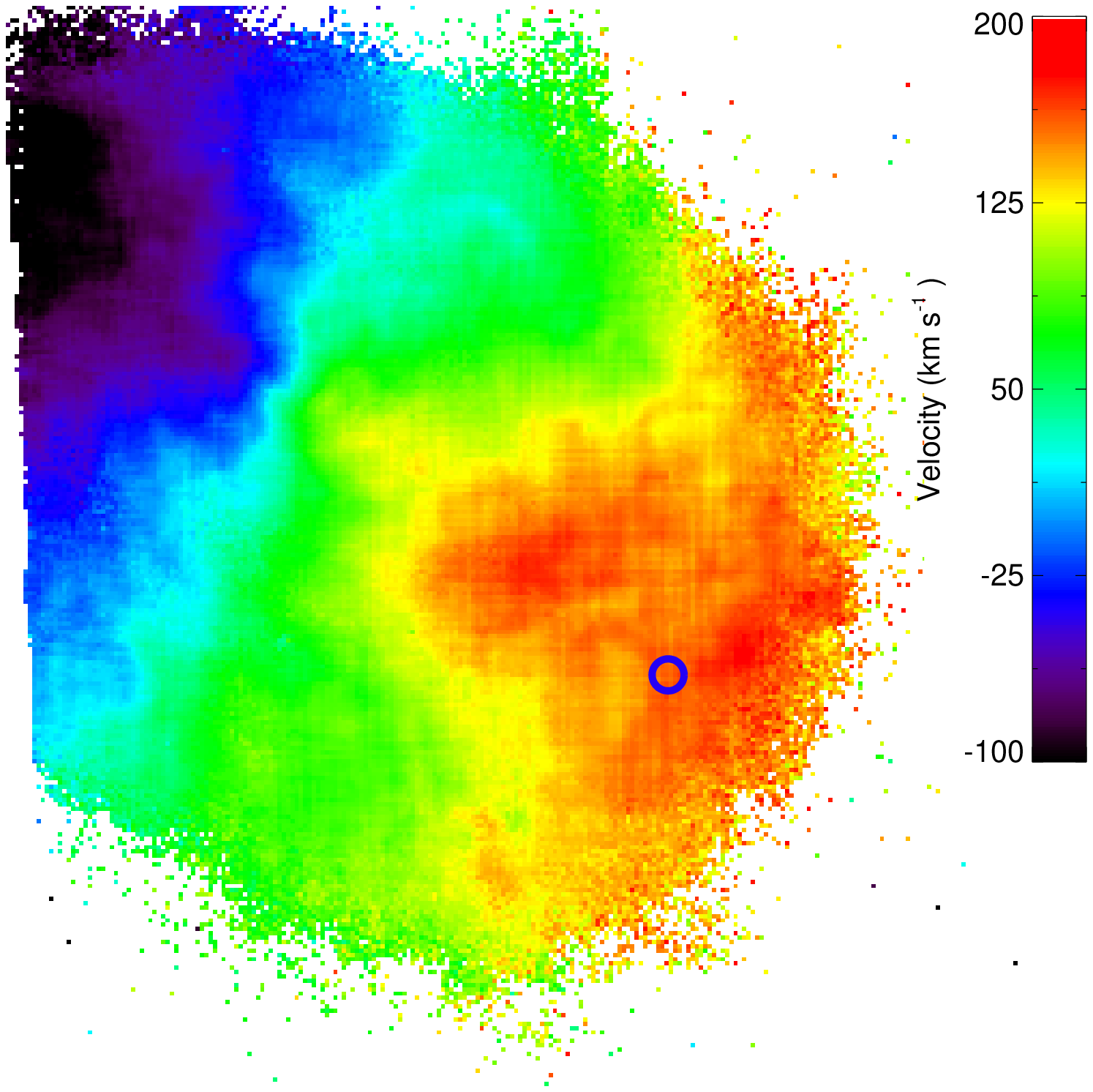} \\
\end{tabular}
\end{center}
\caption{MUSE maps: H$\alpha$ flux, equivalent width, SFR density from H$\alpha$ flux, stellar mass density from $H$-band, specific SFR, and velocity field.
The position of {\sn} is indicated by the blue or red circle. The images are $50\arcsec\times50\arcsec$ (not the entire MUSE coverage). 
North is up and east is left.
White indicates values above the maximum value in the colour bars.
The velocities are relative to the systemic velocity of $2961\,\kms$.
}
\label{fig:muse}
\end{figure*}

\begin{figure*}
\begin{center}
\begin{tabular}{cc}
\includegraphics[width=0.45\textwidth]{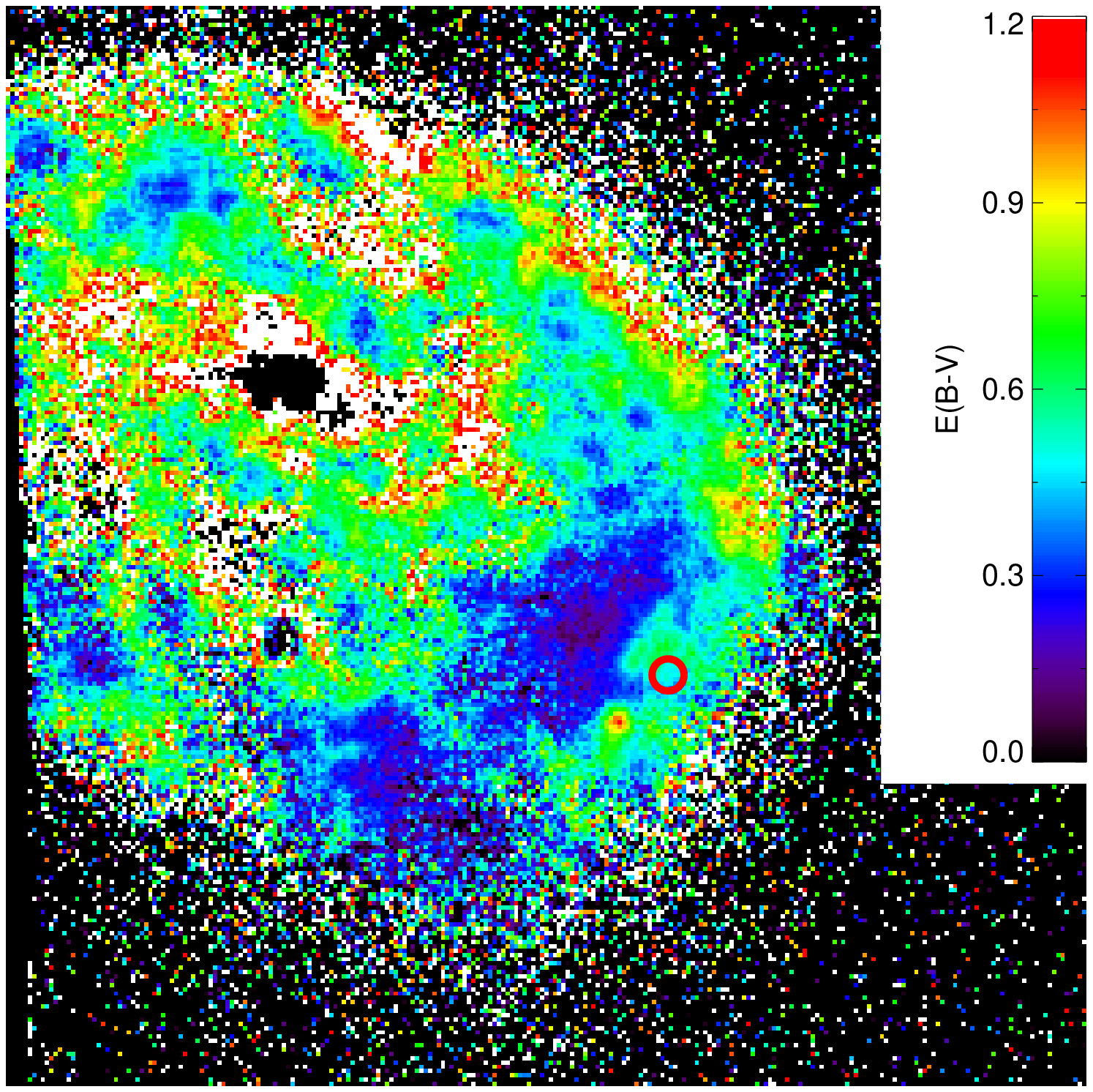} &
\includegraphics[width=0.45\textwidth]{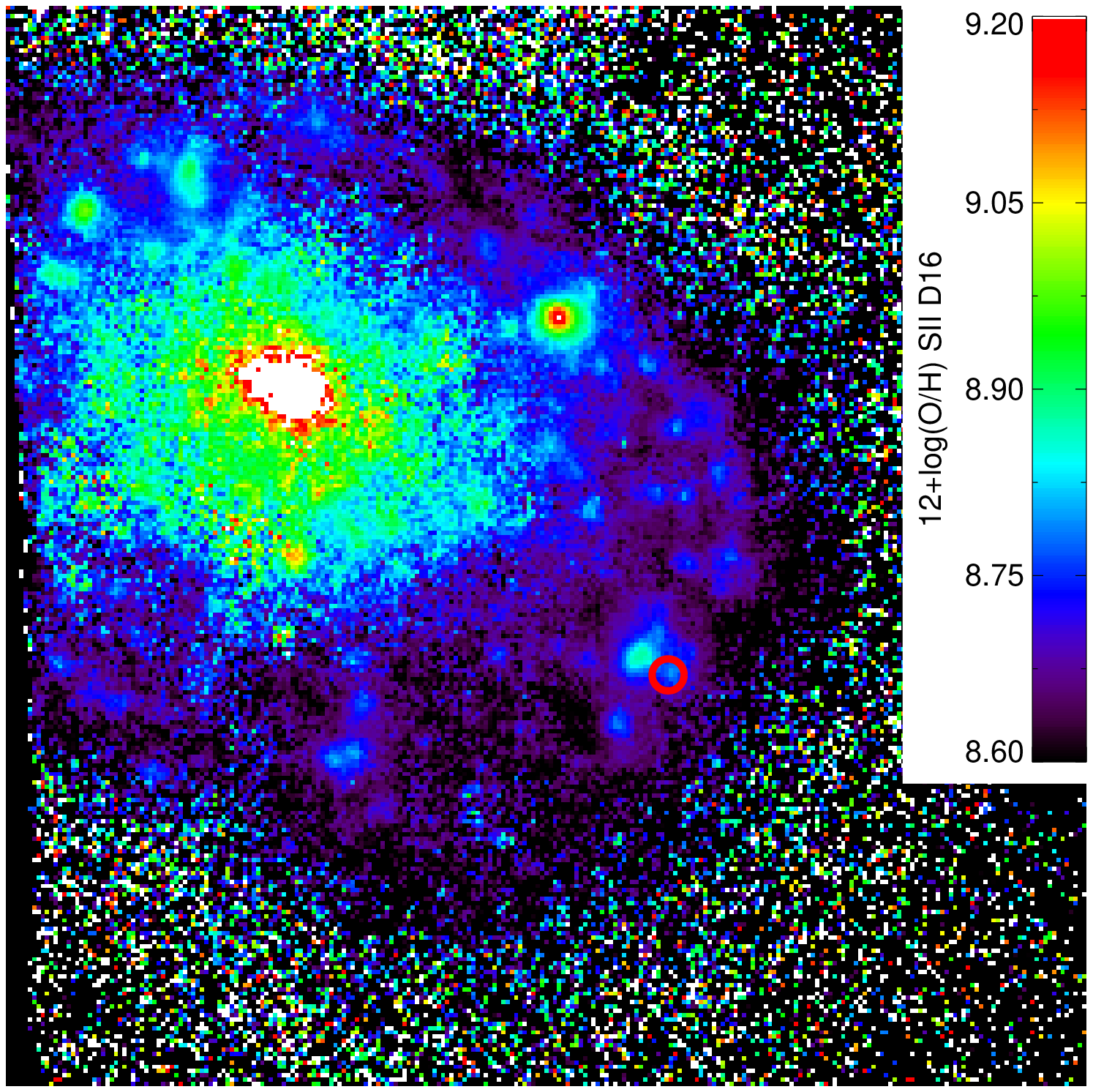} \\ 
\includegraphics[width=0.45\textwidth]{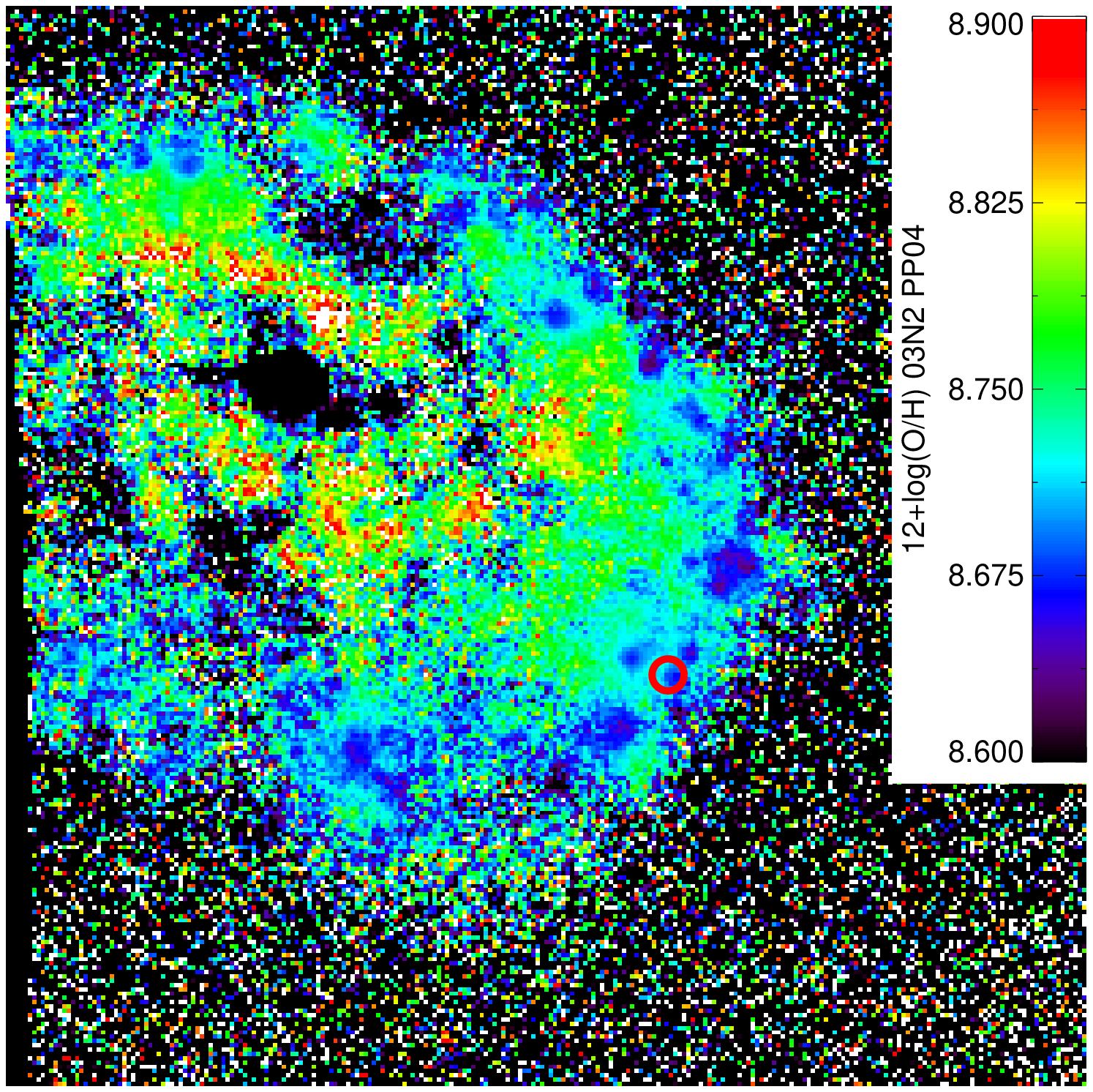} &
\includegraphics[width=0.45\textwidth]{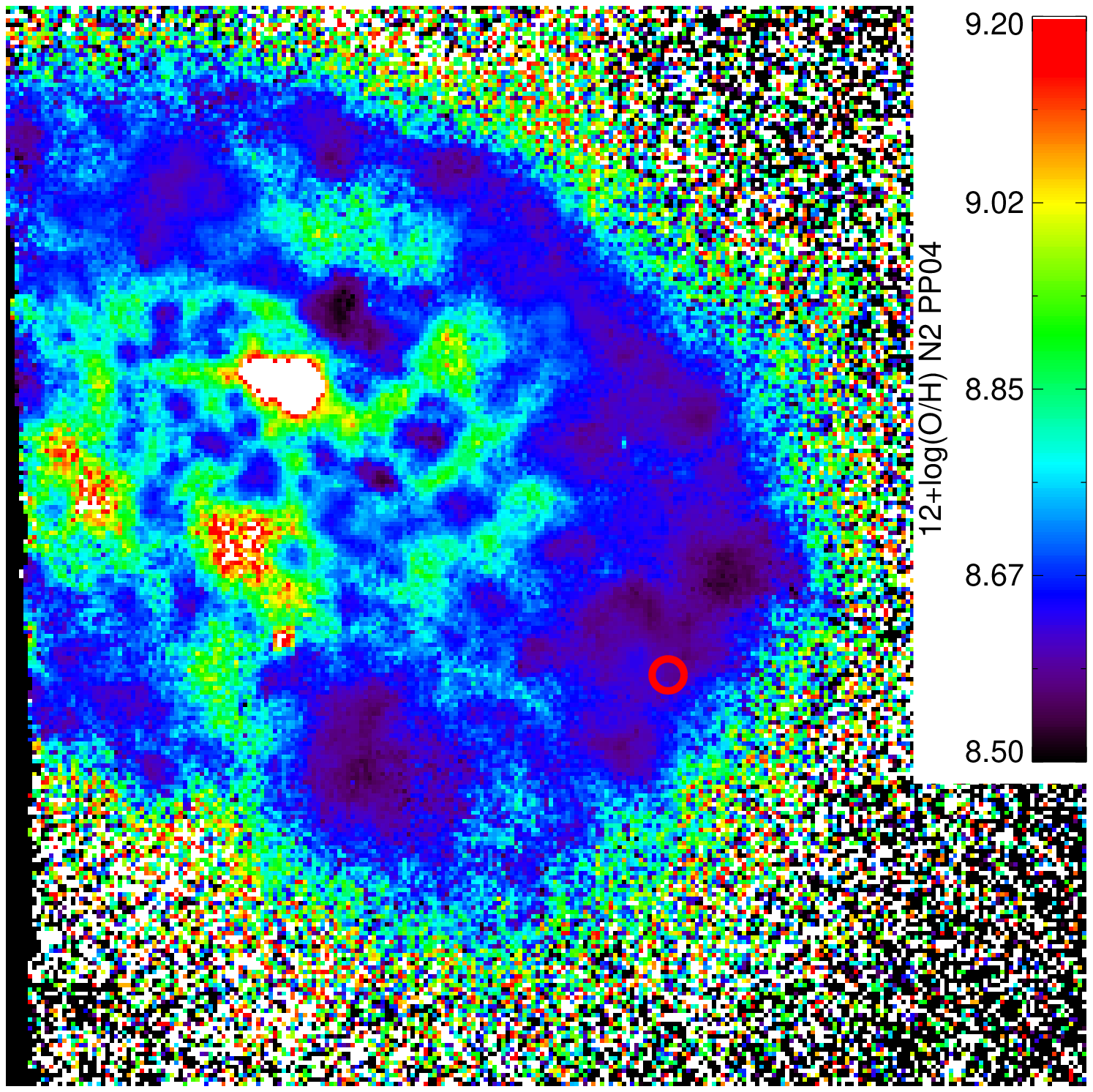}  \\
\end{tabular}   
\end{center}
\caption{Same as for Fig.~\ref{fig:muse}, but for extinction  and the metallicity indicators based on [SII], [NII], and H$\alpha$ fluxes \citep{dopita16}, [OIII], [NII], H$\alpha$, and H$\beta$ lines \citep{pettini04}, and on  [NII] and H$\alpha$ lines \citep{pettini04}.
}
\label{fig:muse1}
\end{figure*}
        
\begin{figure*}
\begin{center}
\includegraphics[width=0.9\textwidth]{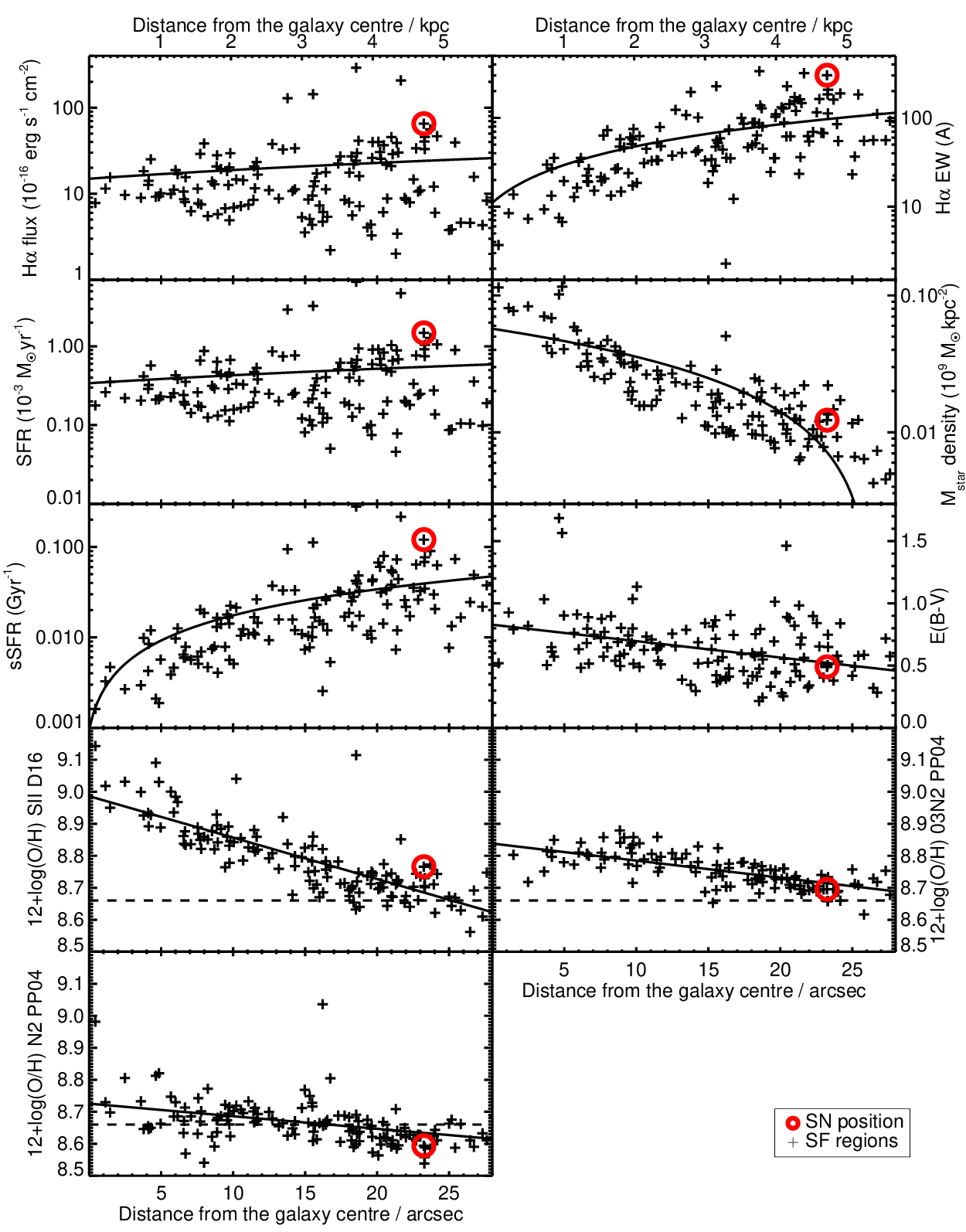}     
\end{center}
\caption{Properties of H$\alpha$-selected star-forming regions as a function of deprojected distance  from the galaxy centre (Sect.~\ref{sec:optifs}): H$\alpha$ flux, equivalent width, SFR from H$\alpha$ flux, stellar mass density, specific SFR, extinction, and three metallicity measurements based on [SII], [NII], and H$\alpha$ fluxes \citep{dopita16}, [OIII], [NII], H$\alpha$, and H$\beta$ lines, and just [NII] and H$\alpha$ lines \citep{pettini04}. The region in which {\sn} exploded is indicated by red circles. The linear fits to the data (Table~\ref{tab:distfit}) are shown as solid lines. The  solar metallicity of $\metoh\sim8.66$ \citep{asplund04} is marked as a dashed line.
}
\label{fig:musedist}
\end{figure*}       

Figure~\ref{fig:contmap} shows the 1.4\,Ghz continuum image from \citet{condon96}. The emission is lopsided and the peak of the emission is close to the position of {\sn}.

Based on the MUSE observations, the distribution of H$\alpha$ flux, equivalent width (EW), SFR, and the velocity field is shown in Fig.~\ref{fig:muse} and the distribution of  dust extinction and metallicity is shown in Fig.~\ref{fig:muse1}.
We derived SFRs of each spaxel from the H$\alpha$ fluxes using the calibration of \citet{kennicutt} with the \citet{chabrier03} IMF.
The dust extinction was derived from the Balmer decrement.
We made three metallicity measurements based on [SII], [NII], and H$\alpha$ fluxes \citep[][used in all analysis unless stated otherwise]{dopita16}, [OIII], [NII], H$\alpha$, and H$\beta$ lines (O3N2),  and just [NII] and H$\alpha$ (N2) lines \citep{pettini04}.

The properties of H$\alpha$-detected star-forming regions were extracted in apertures with radius of 0.5\arcsec ($\sim100\,$pc) from these maps, shown in Fig.~\ref{fig:musedist} as a function of a deprojected galactocentric distance (and in Fig.~\ref{fig:musedistnodeproj} using measured instead of deprojected distances), and listed in Table~\ref{tab:values}. 
They were visually selected in the H$\alpha$ map down to approximately $10^{-16}\,\mbox{erg}\,\mbox{s}^{-1}\,\mbox{cm}^{-2}$. This corresponds to the H$\alpha$ luminosity of $\sim2\times10^{37}\,\mbox{erg}\,\mbox{s}^{-1}$ at the redshift of {\ngc}, which is comparable to the luminosity of \ion{H}{ii} regions in the Milky Way and nearby galaxies \citep[e.g.][]{crowther13}.
The first row in Table~\ref{tab:values} is the {\sn} position and the second is the centre of the galaxy. The last row shows the sum of the individual regions for extensive properties (H$\alpha$ flux and SFR) and the average for the intensive properties (equivalent width, extinction, and metallicities).  
The parameters of the linear fit of the properties as a function of distance from the galaxy centre are presented in Table~\ref{tab:distfit}.
The SN region is one of the most star-forming regions within its host in terms of both SFR and sSFR.

The metallicities at the SN position 
and the nearest bright star-forming region (the fifth row)  are $\sim0.2$--$0.4$\,dex lower than the value of $\metoh\sim8.96\pm0.10$ reported by \citet{levesque10d}. This is because they used the [NII]/[OIII] method of  \citet{kewley02}, which was shown to result in systematically higher metallicities than the methods we employed \citep{kewley08}.
Indeed using the fluxes reported by  \citet{levesque10d} for the SN region, we obtained  $\metoh\sim8.68$ and $\sim8.63$, for the O3N2 and N2 methods of \citet{pettini04}, consistent with our results.
The velocity field derived from the H$\alpha$ line (Fig.~\ref{fig:muse}) is typical for a rotating disk galaxy.

\section{Discussion}
\label{sec:discussion}

In summary, {\ngc} has an enhanced SFR given its stellar mass (close to the starburst regime above the main sequence, Sect.~\ref{sec:stell}), low atomic and molecular gas masses given its SFR (Sect.~\ref{sec:gas}), and the SN region is one of the most star-forming regions  (Figs.~\ref{fig:musedist} and \ref{fig:contmap}). The atomic gas distribution is not centred on the optical galaxy centre, but instead around a third of atomic gas resides in the region close to the SN position (Fig.~\ref{fig:himap}).  These properties are similar to the {\hi} concentrations close to the GRB positions \citep{michalowski15hi, arabsalmani15b} and to the claimed molecular deficiency of GRB hosts \citep{hatsukade14,stanway15,michalowski16}. This was interpreted in \citet{michalowski15hi,michalowski16} as an indication that a very recent inflow of metal-poor atomic gas is responsible for enhanced SFRs, and, in turn, for the birth of the GRB progenitors. This is likely the case for {\sn}. The fact that we executed the high-resolution ATCA {\hi} observations of a relativistic SN host and obtained a similar unusual distribution suggests that both explosion classes prefer similar environments. This needs to be tested with a larger sample of SN hosts observed at {\hi}.
The recent inflow of gas for {\ngc} is also supported by the 
relatively low 
metallicity measured
in the southwestern part of the galaxy (Fig.~\ref{fig:muse1}), close to the {\hi} peak.

The {\hi} velocity field also points at the external origin of at least some of the atomic gas. The {\hi} velocity field (Fig.~\ref{fig:himap}) is not consistent with a rotating disk, as opposed to the H$\alpha$ velocity field (Fig.~\ref{fig:muse}). Moreover, the values of velocities derived from the {\hi} and H$\alpha$ lines are not consistent at the same positions. For example, close to the SN position the H$\alpha$ line gives $\sim200\,\kms$, whereas the {\hi} line results in $\sim50\,\kms$.

{\sn} exploded close to the region with the highest SFR density and the lowest age, as evident from the high H$\alpha$ EW (Figs.~\ref{fig:muse} and \ref{fig:musedist}), similarly to other SN Ib/c \citep{galbany14} and SN II \citep{galbany16b}. Following \citet{kuncarayakti13, kuncarayakti13b} we converted the H$\alpha$ EW of the SN site of   $\sim300\,${\AA} to the stellar population age of  $\sim5.5$\,Myr (assuming instantaneous burst and standard \citealt{salpeter} IMF) by comparing to single stellar population models from Starburst99  \citep{leitherer99}. 
This timescale corresponds to the lifetime of a $\sim36\,\msun$ massive star. Although it is not straightforward to infer this as the initial mass of the {\sn} progenitor, the fact that such a young age is observed at the explosion site supports the view that the progenitor may have been one of such massive stars.
This also means that in the scenario of the gas inflow presented above, it must have begun only several Myr ago, consistent with the timescale presented in \citet{michalowski16} for a GRB host.

The metallicity of the site of {\sn} ($\metoh\sim8.77$ or $\sim1.3$ solar using the calibration of \citealt{dopita16}; Table~\ref{tab:values}) is close to the highest values found for other SN Ib/c \citep{thone09,leloudas11,kuncarayakti13,kuncarayakti18,galbany16} and II-L \citep{kuncarayakti13b}. On the other hand, GRBs are usually found in environments with much lower metallicities  \citep{sollerman05,christensen08,modjaz08,thone08,thone14,han10,levesque10c,levesque11, kruhler15,kruhler17,schulze15,japelj16,izzo17,vergani17}. However, there is a growing sample of  GRBs in solar or super-solar environments 
\citep{prochaska09,levesque10b,kruhler12,savaglio12,elliott13,schulze14,hashimoto15,schady15,stanway15b,michalowski18grb}, which can be explained by recent overcoming of the observational bias against dust, resulting in the discovery of massive and metal-rich hosts \citep{hjorth12,perley15,perley16,perley16b}.
This metallicity information may mean that relativistic explosions signal a recent inflow of gas (and subsequent star formation), and their type (GRBs or SNe) is determined by 
 either 
{\it i)} the metallicity of the inflowing gas, so that metal-poor gas results in a GRB explosion and metal-rich gas%
 \footnote{This would either be a minor merger with an evolved galaxy or gas coming from nearby galaxies, or previously ejected gas falling back. Below is stated the evidence of the existence of a galaxy group close to {\ngc}.}
in a relativistic SN explosion without an accompanying GRB  (see also \citealt{modjaz11,leloudas10,leloudas11}), or {\it ii)} by 
the efficiency of gas mixing (efficient mixing for SN hosts leading to quick disappearance of metal-poor regions), or {\it iii)} by the type of the galaxy (more metal-rich galaxies would result in only a small fraction of star formation being fuelled by metal-poor gas).  

\citet{stott14} interpreted flat metallicity gradients as a sign of a recent inflow of metal-poor gas, because these gradients correlate with sSFR and the distance above the main sequence (their Figs.~3 and 4). For sSFR and $\mbox{SFR}/\mbox{SFR}_{\rm MS}$ of {\ngc,} their relation predicts the metallicity gradient of $\sim-0.02$ and $-0.01\,\mbox{dex}\,\mbox{kpc}^{-1}$, respectively, using the N2 calibrator.
This is similar to the $-0.019\,\mbox{dex}\,\mbox{kpc}^{-1}$ measured for {\ngc} using this calibrator (Table~\ref{tab:distfit}). 
Moreover, the gradients of $-0.06$ and $-0.03\,\mbox{dex}\,\mbox{kpc}^{-1}$ for the hosts of GRB\,980425 \citep{kruhler17} and 060505 \citep{thone08}, respectively, are also consistent with the trends of \citet{stott14}, using the sSFR values tabulated in \citet{michalowski15hi}. This therefore needs to be investigated with a larger sample of SN and GRB hosts. If these galaxies turn out to have steeper metallicity gradients, this would support the scenario that the gas inflow is of higher metallicity or smaller in quantity, so it does not flatten the metallicity gradient. 

Finally, the distributions of the H$\alpha$ emission (Fig.~\ref{fig:muse}), of the optical emission, and especially the radio continuum emission (Fig.~\ref{fig:contmap}) 
of {\ngc} are clearly lopsided, with the western (right) side more pronounced and rich in star-forming regions. Such asymmetry may be a sign of interaction \citep{sancisi08,rasmussen06}, so we  looked at the large-scale environment of {\ngc} using the NASA/IPAC Extragalactic Database (NED). We found a galaxy group designated 0509 in \citet{tully08} at coordinates 10:27:10.4, $-$39:52:58,~ $\sim51'$ or $\sim600$\,kpc west in projection from {\ngc} with $z=0.009493$ (similar to the redshift of {\ngc}, shifted only by $\sim100$\,\kms). The distance of 600\,kpc may be too high to influence {\ngc} that strongly (e.g.~the clear sign of interaction reported by \citealt{rasmussen06} concerns a galaxy $\sim70$\,kpc from the group). However the existence of the galaxy group in the vicinity of {\ngc} indicates that indeed there  should be a significant supply of ambient intergalactic gas available for inflow onto this galaxy.

The caveat of this work is that {\sn} was discovered in a galaxy-targeted survey. Such surveys were shown to result on average in higher metallicities and masses than un-targeted surveys \citep{sanders12}. Therefore analysis of the gas content of a larger sample of relativistic SN hosts from both targeted and un-targeted surveys is needed. We note, however, that hosts of broad-line Ic SN from both targeted and un-targeted surveys do include objects with metallicities around solar \citep{sanders12}, similar to {\ngc}.

\section{Conclusions}
\label{sec:conclusion}

We obtained 21\,cm hydrogen line ({\hi}) and optical integral field unit spectroscopy observations of {\ngc}, the host galaxy of the relativistic {\sn}. 
This is the first time the atomic gas properties of a relativistic SN host have been provided and the first time resolved 21\,cm-hydrogen-line ({\hi}) information is analysed for the host of an SN of any type in the context of the SN position.
The atomic gas distribution of {\ngc} is not centred on the optical galaxy centre, but instead around a third of the atomic gas resides in the region close to the SN position. This galaxy has a few times lower atomic and molecular gas masses than predicted from its star formation rate (SFR). Its specific star formation rate ($\mbox{sSFR}\equiv\mbox{SFR}/\mstar$) is approximately two to three times higher than the main-sequence value, placing it at the higher end of the main sequence towards starburst galaxies. {\sn} exploded close to the region with the highest SFR density and the lowest age, as evident from a high H$\alpha$ EW, corresponding to the age of the stellar population of $\sim5.5$\,Myr. 
Assuming this timescale was the lifetime of the progenitor star, its initial mass would have been close to $\sim36\,\msun$.
The gas properties of {\ngc} are consistent with a recent inflow of gas from the intergalactic medium, which explains the concentration of atomic gas close to the SN position and the enhanced SFR. Super-solar metallicity at the position of the SN (unlike for most GRBs) may mean that relativistic explosions signal a recent inflow of gas (and subsequent star formation), and their type (GRBs or SNe) is determined by either {\it i)} 
{\it } the metallicity of the inflowing gas, so that metal-poor gas results in a GRB explosion and metal-rich gas 
(for example a minor merger with an evolved galaxy or re-accretion of expelled gas)
in a relativistic SN explosion without an accompanying GRB,
{\it ii)} the efficiency of gas mixing (efficient mixing for SN hosts leading to quick disappearance of metal-poor regions), or {\it iii)}  the type of the galaxy (more metal-rich galaxies would result in only a small fraction of star formation being fuelled by metal-poor gas). 

\begin{acknowledgements}

We thank Joanna Baradziej for help to improve this paper, and Giuliano Pignata, Carlos Contreras, and Maximilian Stritzinger for sharing the $H$-band image.

M.J.M.~acknowledges the support of 
the National Science Centre, Poland, through the POLONEZ grant 2015/19/P/ST9/04010;
and the UK Science and Technology Facilities Council;
this project has received funding from the European Union's Horizon 2020 research and innovation programme under the Marie Sk{\l}odowska-Curie grant agreement No. 665778.
J.H. was supported by a VILLUM FONDEN Investigator grant (project number 16599).
L.G. was supported in part by the US National Science Foundation under Grant AST-1311862.
L.K.H.~acknowledges funding from the INAF PRIN-SKA program 1.05.01.88.04.
A.d.U.P.~acknowledges support from the European Commission (FP7-PEOPLE-2012-CIG 322307) and from the Spanish project AYA2012-39362-C02-02.
S.D.V.~is supported by the French National Research Agency (ANR) under contract ANR-16-CE31-0003 BEaPro.

The Australia Telescope Compact Array is part of the Australia Telescope National Facility, which is funded by the Commonwealth of Australia for operation as a National Facility managed by CSIRO. 
Based on observations collected at the European Organisation for Astronomical Research in the Southern Hemisphere under ESO programme(s) 095.D-0172(A). 
This publication makes use of data products from the Wide-field Infrared Survey Explorer, which is a joint project of the University of California, Los Angeles, and the Jet Propulsion Laboratory/California Institute of Technology, funded by the National Aeronautics and Space Administration. 
We acknowledge the usage of the HyperLeda database (\url{http://leda.univ-lyon1.fr}).
This research has made use of 
the GHostS database (\urltt{http://www.grbhosts.org}), which is partly funded by Spitzer/NASA grant RSA Agreement No. 1287913; 
the NASA/IPAC Extragalactic Database (NED), which is operated by the Jet Propulsion Laboratory, California Institute of Technology, under contract with the National Aeronautics and Space Administration;
SAOImage DS9, developed by Smithsonian Astrophysical Observatory \citep{ds9};
and  NASA's Astrophysics Data System Bibliographic Services.

\end{acknowledgements}



\appendix
\section{Long tables and additional figures}

\begin{figure}
\begin{center}
\includegraphics[width=0.45\textwidth,clip]{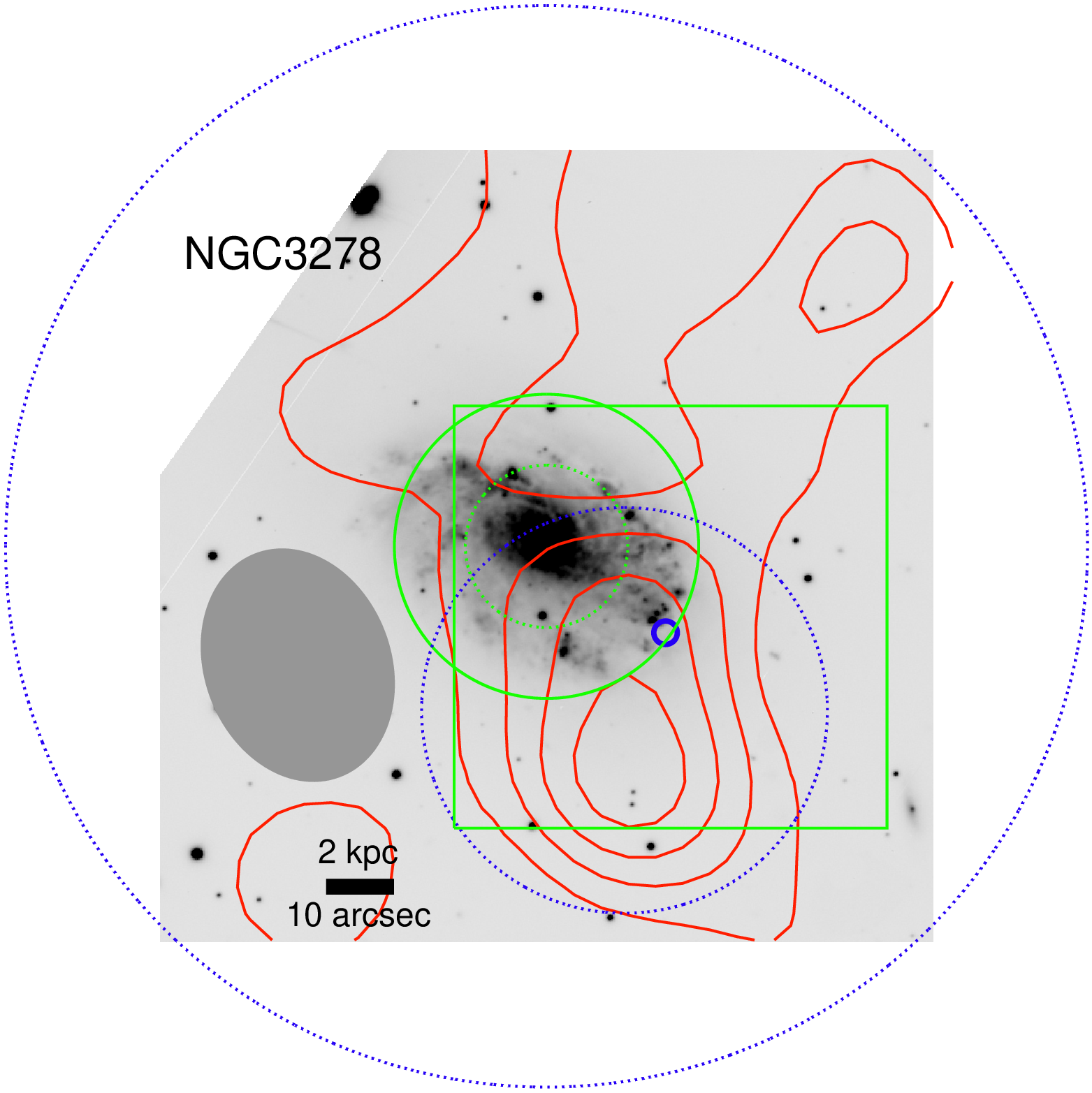} 
\end{center}
\caption{Similar to Fig.~\ref{fig:himap} with additional indication of the spatial coverage of other observations.  Green solid and dotted circles correspond to the FWHM of the CO(1-0) and CO(2-1) observations from \citet{albrecht07}. The green square shows the position of the MUSE observations. Blue dotted circles indicate the apertures used to extract the {\hi} spectrum of the entire galaxy (larger than the image) and the {\hi} peak.
}
\label{fig:himapwithbox}
\end{figure}

\begin{figure*}
\begin{center}
\includegraphics[width=0.9\textwidth]{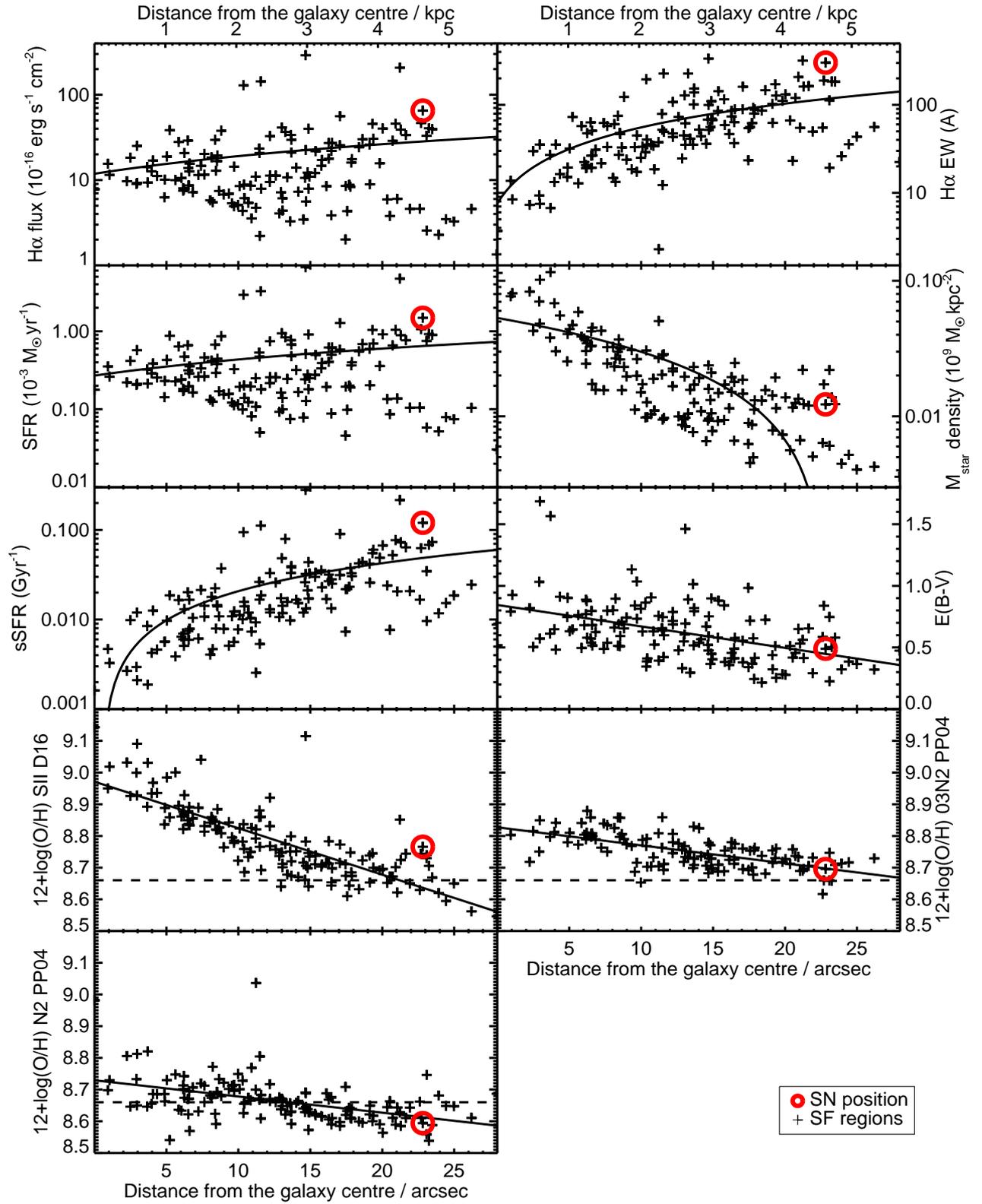}     
\end{center}
\caption{Same as Fig.~\ref{fig:musedist} but using measured, not deprojected, distance from the galaxy centre.
}
\label{fig:musedistnodeproj}
\end{figure*}       

\clearpage
\onecolumn
\begin{longtable}{ccrcrrrrccccccc}
\caption{Properties of star-forming regions derived from the MUSE data.\label{tab:values}}
\\
\hline\hline
RA & Dec & \multicolumn{1}{c}{x} & \multicolumn{1}{c}{y} & \multicolumn{1}{c}{dist$_{\rm centre}$} & \multicolumn{1}{c}{dist$_{\rm centre}^{\rm deproj.}$} & \multicolumn{1}{c}{H$\alpha$} & \multicolumn{1}{c}{H$\alpha$ EW} & SFR & $\mstar$ & sSFR & E(B-V) & \multicolumn{3}{c}{$\metoh$}    \\
\cline{13-15}
 (deg) & (deg) & \multicolumn{1}{c}{(pix)} & \multicolumn{1}{c}{(pix)} & \multicolumn{1}{c}{($\arcsec$)} & \multicolumn{1}{c}{($\arcsec$)} &   & \multicolumn{1}{c}{(\r{A})} & & & (Gyr$^{-1}$) & (mag) & D16 & 03N2 & N2 \\
(1)    & (2) & \multicolumn{1}{c}{(3)} & \multicolumn{1}{c}{(4)} & \multicolumn{1}{c}{(5)} & \multicolumn{1}{c}{(6)} & \multicolumn{1}{c}{(7)} & \multicolumn{1}{c}{(8)} & (9) & (10) & (11) & (12) & (13) & (14) & (15)    \\
\hline
\endfirsthead
\caption{Continued.}\\
\hline\hline
RA & Dec & \multicolumn{1}{c}{x} & \multicolumn{1}{c}{y} & \multicolumn{1}{c}{dist$_{\rm centre}$} & \multicolumn{1}{c}{dist$_{\rm centre}^{\rm deproj.}$} & \multicolumn{1}{c}{H$\alpha$} & \multicolumn{1}{c}{H$\alpha$ EW} & SFR & $\mstar$ & sSFR & E(B-V) & \multicolumn{3}{c}{$\metoh$}    \\
\cline{13-15}
 (deg) & (deg) & \multicolumn{1}{c}{(pix)} & \multicolumn{1}{c}{(pix)} & \multicolumn{1}{c}{($\arcsec$)} & \multicolumn{1}{c}{($\arcsec$)} &   & \multicolumn{1}{c}{(\r{A})} & & & (Gyr$^{-1}$) & (mag) & D16 & 03N2 & N2 \\
(1)    & (2) & \multicolumn{1}{c}{(3)} & \multicolumn{1}{c}{(4)} & \multicolumn{1}{c}{(5)} & \multicolumn{1}{c}{(6)} & \multicolumn{1}{c}{(7)} & \multicolumn{1}{c}{(8)} & (9) & (10) & (11) & (12) & (13) & (14) & (15)    \\
\hline
\endhead
\hline
\endfoot
157.8911250 & -39.9583611 & 156.19 & 148.15 & 22.82 & 23.25 & 65 & 302.3 & \phantom{1}1.484 & 0.012 & 0.120 & 0.49 & 8.77 & 8.70 & 8.59 \\
157.8980255 & -39.9548683 & 60.98 & 211.01 & 0.00 & 0.44 & 8 & 3.7 & \phantom{1}0.178 & 0.114 & 0.002 & 0.52 & 9.14 & $\cdots$ & 8.98 \\
157.8992462 & -39.9521484 & 44.13 & 259.97 & 10.35 & 13.78 & 129 & 195.0 & \phantom{1}2.933 & 0.031 & 0.095 & 0.35 & 8.84 & 8.77 & 8.63 \\
157.8929138 & -39.9537468 & 131.51 & 231.20 & 14.67 & 18.54 & 291 & 335.3 & \phantom{1}6.619 & 0.024 & 0.279 & 0.47 & 9.11 & 8.68 & 8.69 \\
157.8916321 & -39.9581451 & 149.19 & 152.03 & 21.22 & 21.65 & 207 & 318.7 & \phantom{1}4.722 & 0.022 & 0.216 & 0.56 & 8.85 & 8.69 & 8.63 \\
157.8973999 & -39.9568672 & 69.61 & 175.03 & 7.40 & 10.22 & 15 & 62.3 & \phantom{1}0.334 & 0.016 & 0.021 & 0.49 & 9.04 & 8.84 & 8.72 \\
157.8970337 & -39.9512634 & 74.66 & 275.90 & 13.26 & 20.46 & 37 & 226.9 & \phantom{1}0.833 & 0.010 & 0.079 & 0.89 & 8.77 & 8.73 & 8.62 \\
157.8949280 & -39.9519005 & 103.72 & 264.44 & 13.68 & 20.99 & 21 & 153.4 & \phantom{1}0.478 & 0.013 & 0.038 & 0.82 & 8.67 & 8.71 & 8.59 \\
157.8941650 & -39.9527893 & 114.24 & 248.44 & 13.02 & 18.70 & 41 & 128.2 & \phantom{1}0.940 & 0.020 & 0.048 & 0.60 & 8.75 & 8.77 & 8.67 \\
157.8968048 & -39.9537582 & 77.82 & 231.00 & 5.23 & 7.99 & 38 & 72.9 & \phantom{1}0.874 & 0.047 & 0.019 & 0.65 & 8.86 & $\cdots$ & 8.54 \\
157.8952942 & -39.9537697 & 98.66 & 230.79 & 8.51 & 11.71 & 17 & 48.0 & \phantom{1}0.380 & 0.026 & 0.014 & 0.51 & 8.82 & 8.80 & 8.69 \\
157.8992767 & -39.9518051 & 43.71 & 266.15 & 11.56 & 15.55 & 143 & 227.1 & \phantom{1}3.261 & 0.029 & 0.112 & 0.42 & 8.86 & 8.69 & 8.61 \\
157.9010468 & -39.9523964 & 19.29 & 255.50 & 12.19 & 13.46 & 33 & 107.8 & \phantom{1}0.744 & 0.022 & 0.033 & 0.79 & 8.92 & 8.81 & 8.66 \\
157.8983459 & -39.9524345 & 56.55 & 254.82 & 8.81 & 12.73 & 38 & 123.2 & \phantom{1}0.861 & 0.023 & 0.037 & 0.65 & 8.81 & 8.79 & 8.68 \\
157.8983917 & -39.9531555 & 55.92 & 241.84 & 6.25 & 8.86 & 21 & 47.6 & \phantom{1}0.472 & 0.036 & 0.013 & 0.58 & 8.93 & 8.88 & 8.67 \\
157.8978119 & -39.9532509 & 63.92 & 240.13 & 5.85 & 8.88 & 28 & 57.4 & \phantom{1}0.637 & 0.039 & 0.016 & 0.50 & 8.90 & 8.84 & 8.64 \\
157.8957825 & -39.9526939 & 91.93 & 250.15 & 9.98 & 15.32 & 5 & 25.0 & \phantom{1}0.116 & 0.015 & 0.008 & 0.81 & 8.72 & 8.65 & 8.75 \\
157.8932953 & -39.9524117 & 126.24 & 255.24 & 15.77 & 22.53 & 15 & 115.5 & \phantom{1}0.335 & 0.009 & 0.035 & 0.84 & 8.70 & 8.69 & 8.62 \\
157.8904419 & -39.9548607 & 165.62 & 211.16 & 20.93 & 23.32 & 46 & 207.5 & \phantom{1}1.043 & 0.013 & 0.077 & 0.65 & 8.70 & 8.73 & 8.59 \\
157.8929443 & -39.9531212 & 131.09 & 242.47 & 15.37 & 20.64 & 22 & 143.0 & \phantom{1}0.495 & 0.015 & 0.033 & 0.88 & 8.74 & 8.71 & 8.62 \\
157.8931427 & -39.9554749 & 128.35 & 200.10 & 13.65 & 14.29 & 34 & 101.1 & \phantom{1}0.768 & 0.023 & 0.033 & 0.52 & 8.80 & 8.83 & 8.63 \\
157.8910370 & -39.9551773 & 157.41 & 205.46 & 19.32 & 20.99 & 46 & 171.7 & \phantom{1}1.038 & 0.019 & 0.055 & 0.51 & 8.74 & 8.73 & 8.60 \\
157.8902740 & -39.9557800 & 167.93 & 194.61 & 21.64 & 22.71 & 34 & 163.4 & \phantom{1}0.766 & 0.012 & 0.064 & 0.65 & 8.74 & 8.70 & 8.61 \\
157.8978424 & -39.9512405 & 63.50 & 276.31 & 13.07 & 19.57 & 4 & 35.2 & \phantom{1}0.100 & 0.008 & 0.012 & 0.69 & 8.71 & 8.73 & 8.65 \\
157.8981781 & -39.9508362 & 58.87 & 283.59 & 14.52 & 21.40 & 3 & 36.1 & \phantom{1}0.078 & 0.007 & 0.012 & 0.67 & 8.72 & 8.71 & 8.64 \\
157.8985443 & -39.9511108 & 53.81 & 278.65 & 13.60 & 19.62 & 3 & 35.3 & \phantom{1}0.075 & 0.008 & 0.010 & 1.01 & 8.75 & 8.70 & 8.66 \\
157.9005280 & -39.9512634 & 26.44 & 275.90 & 14.70 & 18.27 & 6 & 51.9 & \phantom{1}0.128 & 0.010 & 0.013 & 0.75 & 8.71 & 8.73 & 8.65 \\
157.9012146 & -39.9506836 & 16.97 & 286.33 & 17.45 & 21.30 & 2 & 23.4 & \phantom{1}0.046 & 0.006 & 0.007 & 0.98 & 8.71 & $\cdots$ & 8.71 \\
157.9019928 & -39.9514503 & 6.23 & 272.53 & 16.47 & 18.37 & 22 & 63.5 & \phantom{1}0.500 & 0.018 & 0.028 & 0.41 & 8.65 & 8.73 & 8.59 \\
157.9004364 & -39.9529076 & 27.71 & 246.30 & 9.70 & 10.69 & 14 & 44.9 & \phantom{1}0.326 & 0.026 & 0.012 & 0.73 & 8.79 & $\cdots$ & 8.70 \\
157.9015198 & -39.9531555 & 12.76 & 241.84 & 11.45 & 11.64 & 23 & 78.9 & \phantom{1}0.533 & 0.020 & 0.026 & 0.66 & 8.86 & 8.76 & 8.67 \\
157.8997955 & -39.9529762 & 36.55 & 245.07 & 8.38 & 9.81 & 30 & 75.1 & \phantom{1}0.673 & 0.030 & 0.022 & 0.51 & 8.85 & 8.86 & 8.62 \\
157.8991852 & -39.9529419 & 44.97 & 245.69 & 7.64 & 9.74 & 20 & 57.6 & \phantom{1}0.461 & 0.032 & 0.014 & 0.80 & 8.83 & 8.83 & 8.66 \\
157.8995972 & -39.9536972 & 39.29 & 232.09 & 6.05 & 6.56 & 11 & 22.4 & \phantom{1}0.241 & 0.037 & 0.006 & 0.68 & 8.86 & 8.80 & 8.63 \\
157.8984833 & -39.9537125 & 54.66 & 231.82 & 4.35 & 5.88 & 11 & 16.3 & \phantom{1}0.254 & 0.055 & 0.005 & 0.91 & 8.94 & $\cdots$ & 8.69 \\
157.8978424 & -39.9527512 & 63.50 & 249.12 & 7.64 & 11.51 & 12 & 33.2 & \phantom{1}0.263 & 0.025 & 0.011 & 0.54 & 8.81 & 8.75 & 8.71 \\
157.8991699 & -39.9533806 & 45.19 & 237.79 & 6.22 & 7.58 & 8 & 19.4 & \phantom{1}0.176 & 0.035 & 0.005 & 0.88 & 8.87 & 8.81 & 8.74 \\
157.9002533 & -39.9542427 & 30.24 & 222.27 & 6.55 & 6.59 & 9 & 19.3 & \phantom{1}0.197 & 0.033 & 0.006 & 0.80 & 8.84 & 8.81 & 8.67 \\
157.8995361 & -39.9541588 & 40.13 & 223.78 & 4.89 & 4.97 & 10 & 19.4 & \phantom{1}0.230 & 0.040 & 0.006 & 0.90 & 8.89 & 8.78 & 8.66 \\
157.8985443 & -39.9541626 & 53.82 & 223.72 & 2.92 & 3.60 & 9 & 9.3 & \phantom{1}0.205 & 0.070 & 0.003 & 1.03 & 9.00 & 8.75 & 8.73 \\
157.8975677 & -39.9541245 & 67.29 & 224.40 & 2.96 & 4.63 & 9 & 7.5 & \phantom{1}0.210 & 0.102 & 0.002 & 1.69 & 9.09 & $\cdots$ & 8.81 \\
157.8958130 & -39.9542313 & 91.51 & 222.48 & 6.52 & 8.54 & 14 & 26.4 & \phantom{1}0.315 & 0.042 & 0.007 & 0.54 & 8.88 & 8.80 & 8.62 \\
157.8946533 & -39.9533195 & 107.51 & 238.89 & 10.85 & 15.22 & 9 & 34.1 & \phantom{1}0.195 & 0.019 & 0.010 & 0.63 & 8.75 & 8.73 & 8.67 \\
157.8937988 & -39.9538879 & 119.30 & 228.66 & 12.19 & 15.51 & 9 & 47.4 & \phantom{1}0.216 & 0.017 & 0.013 & 0.72 & 8.83 & 8.75 & 8.72 \\
157.8935699 & -39.9532013 & 122.46 & 241.02 & 13.68 & 18.61 & 30 & 138.2 & \phantom{1}0.674 & 0.017 & 0.039 & 0.79 & 8.72 & 8.73 & 8.63 \\
157.8960419 & -39.9515724 & 88.35 & 270.34 & 13.07 & 20.42 & 6 & 61.6 & \phantom{1}0.139 & 0.010 & 0.014 & 1.46 & 8.70 & $\cdots$ & 8.67 \\
157.8941345 & -39.9520073 & 114.66 & 262.52 & 14.88 & 22.27 & 8 & 69.4 & \phantom{1}0.180 & 0.011 & 0.017 & 0.90 & 8.64 & 8.70 & 8.63 \\
157.8993530 & -39.9548950 & 42.66 & 210.53 & 3.66 & 4.14 & 14 & 27.0 & \phantom{1}0.323 & 0.038 & 0.008 & 0.61 & 8.89 & 8.80 & 8.65 \\
157.8983917 & -39.9547997 & 55.93 & 212.25 & 1.04 & 1.14 & 11 & 8.4 & \phantom{1}0.261 & 0.081 & 0.003 & 0.93 & 9.02 & $\cdots$ & 8.73 \\
157.8927460 & -39.9543610 & 133.82 & 220.15 & 14.68 & 17.30 & 22 & 99.6 & \phantom{1}0.511 & 0.016 & 0.032 & 0.59 & 8.78 & 8.78 & 8.69 \\
157.8950348 & -39.9553680 & 102.24 & 202.02 & 8.45 & 8.71 & 13 & 27.1 & \phantom{1}0.293 & 0.037 & 0.008 & 0.69 & 8.82 & $\cdots$ & 8.59 \\
157.8954468 & -39.9552116 & 96.56 & 204.84 & 7.22 & 7.54 & 11 & 21.7 & \phantom{1}0.261 & 0.043 & 0.006 & 0.72 & 8.89 & 8.76 & 8.63 \\
157.8939667 & -39.9548569 & 116.98 & 211.22 & 11.20 & 12.50 & 11 & 37.6 & \phantom{1}0.250 & 0.027 & 0.009 & 0.76 & 8.79 & 8.81 & 8.70 \\
157.8936310 & -39.9544907 & 121.61 & 217.82 & 12.20 & 14.29 & 11 & 43.2 & \phantom{1}0.257 & 0.021 & 0.012 & 0.69 & 8.77 & 8.78 & 8.67 \\
157.8998718 & -39.9560280 & 35.51 & 190.14 & 6.59 & 9.67 & 13 & 50.8 & \phantom{1}0.287 & 0.017 & 0.017 & 0.68 & 8.89 & 8.81 & 8.69 \\
157.8986359 & -39.9553795 & 52.56 & 201.81 & 2.49 & 3.79 & 18 & 29.8 & \phantom{1}0.416 & 0.043 & 0.010 & 0.50 & 8.93 & 8.81 & 8.65 \\
157.8979340 & -39.9556847 & 62.24 & 196.32 & 2.95 & 4.29 & 25 & 35.4 & \phantom{1}0.572 & 0.048 & 0.012 & 0.57 & 8.93 & 8.85 & 8.65 \\
157.8981476 & -39.9559937 & 59.30 & 190.76 & 4.07 & 6.15 & 17 & 35.3 & \phantom{1}0.386 & 0.030 & 0.013 & 0.48 & 8.97 & 8.84 & 8.69 \\
157.8920441 & -39.9549637 & 143.51 & 209.30 & 16.51 & 18.22 & 27 & 88.9 & \phantom{1}0.617 & 0.025 & 0.025 & 0.58 & 8.70 & 8.77 & 8.61 \\
157.8921509 & -39.9552956 & 142.04 & 203.33 & 16.28 & 17.45 & 27 & 85.7 & \phantom{1}0.618 & 0.019 & 0.032 & 0.46 & 8.71 & 8.78 & 8.61 \\
157.8912659 & -39.9560280 & 154.25 & 190.14 & 19.12 & 19.65 & 26 & 102.7 & \phantom{1}0.593 & 0.014 & 0.042 & 0.44 & 8.74 & 8.75 & 8.62 \\
157.8923950 & -39.9562187 & 138.67 & 186.71 & 16.28 & 16.44 & 18 & 73.8 & \phantom{1}0.405 & 0.015 & 0.027 & 0.41 & 8.78 & 8.76 & 8.67 \\
157.8960876 & -39.9553566 & 87.72 & 202.23 & 5.63 & 5.68 & 10 & 12.8 & \phantom{1}0.227 & 0.059 & 0.004 & 0.76 & 9.00 & 8.77 & 8.75 \\
157.8959961 & -39.9558411 & 88.98 & 193.51 & 6.61 & 6.70 & 16 & 26.6 & \phantom{1}0.355 & 0.045 & 0.008 & 0.52 & 8.83 & 8.80 & 8.57 \\
157.8967285 & -39.9563713 & 78.88 & 183.96 & 6.49 & 7.72 & 29 & 63.2 & \phantom{1}0.660 & 0.033 & 0.020 & 0.57 & 8.85 & 8.86 & 8.67 \\
157.8970337 & -39.9560394 & 74.66 & 189.94 & 5.03 & 6.00 & 19 & 31.9 & \phantom{1}0.429 & 0.044 & 0.010 & 0.63 & 8.98 & 8.81 & 8.73 \\
157.8957520 & -39.9564781 & 92.35 & 182.04 & 8.54 & 9.15 & 19 & 46.5 & \phantom{1}0.442 & 0.032 & 0.014 & 0.59 & 8.88 & 8.86 & 8.69 \\
157.8942413 & -39.9561844 & 113.19 & 187.33 & 11.47 & 11.47 & 21 & 52.1 & \phantom{1}0.472 & 0.028 & 0.017 & 0.64 & 8.88 & 8.86 & 8.67 \\
157.8912964 & -39.9544678 & 153.83 & 218.23 & 18.63 & 21.47 & 30 & 146.8 & \phantom{1}0.685 & 0.016 & 0.044 & 0.72 & 8.72 & 8.68 & 8.60 \\
157.8963470 & -39.9567261 & 84.14 & 177.58 & 8.14 & 9.59 & 18 & 60.3 & \phantom{1}0.419 & 0.023 & 0.018 & 0.66 & 8.86 & 8.82 & 8.67 \\
157.8956909 & -39.9570236 & 93.19 & 172.22 & 10.09 & 11.42 & 9 & 33.5 & \phantom{1}0.194 & 0.017 & 0.011 & 0.63 & 8.82 & 8.79 & 8.70 \\
157.8967285 & -39.9570580 & 78.88 & 171.60 & 8.66 & 11.07 & 7 & 32.8 & \phantom{1}0.170 & 0.016 & 0.011 & 0.71 & 8.83 & 8.81 & 8.73 \\
157.8964081 & -39.9574738 & 83.30 & 164.12 & 10.39 & 13.17 & 9 & 40.4 & \phantom{1}0.201 & 0.013 & 0.016 & 0.39 & 8.76 & 8.79 & 8.65 \\
157.8963470 & -39.9581223 & 84.14 & 152.44 & 12.60 & 16.44 & 11 & 73.0 & \phantom{1}0.256 & 0.010 & 0.026 & 0.48 & 8.76 & 8.78 & 8.66 \\
157.8943176 & -39.9572716 & 112.14 & 167.76 & 13.40 & 14.13 & 11 & 41.3 & \phantom{1}0.245 & 0.016 & 0.016 & 0.30 & 8.68 & 8.75 & 8.65 \\
157.8939056 & -39.9575081 & 117.83 & 163.50 & 14.82 & 15.59 & 12 & 56.0 & \phantom{1}0.278 & 0.013 & 0.021 & 0.37 & 8.70 & 8.77 & 8.64 \\
157.8974609 & -39.9553185 & 68.77 & 202.91 & 2.25 & 2.49 & 10 & 7.3 & \phantom{1}0.220 & 0.083 & 0.003 & 0.82 & 9.03 & 8.72 & 8.81 \\
157.8981171 & -39.9551163 & 59.72 & 206.55 & 0.93 & 1.46 & 16 & 13.7 & \phantom{1}0.353 & 0.077 & 0.005 & 0.79 & 8.95 & 8.80 & 8.70 \\
157.9002380 & -39.9573059 & 30.45 & 167.13 & 10.69 & 16.54 & 5 & 35.0 & \phantom{1}0.123 & 0.010 & 0.012 & 0.91 & 8.73 & 8.70 & 8.67 \\
157.9000549 & -39.9576836 & 32.98 & 160.34 & 11.58 & 18.06 & 8 & 49.4 & \phantom{1}0.180 & 0.010 & 0.019 & 0.60 & 8.69 & 8.71 & 8.68 \\
157.8987885 & -39.9574585 & 50.46 & 164.39 & 9.56 & 14.78 & 5 & 33.1 & \phantom{1}0.122 & 0.011 & 0.011 & 0.84 & 8.80 & 8.69 & 8.71 \\
157.8992615 & -39.9576035 & 43.93 & 161.78 & 10.42 & 16.25 & 6 & 35.5 & \phantom{1}0.140 & 0.010 & 0.014 & 0.41 & 8.69 & 8.70 & 8.66 \\
157.8928070 & -39.9566803 & 132.98 & 178.40 & 15.81 & 15.81 & 17 & 59.0 & \phantom{1}0.391 & 0.017 & 0.023 & 0.38 & 8.71 & 8.75 & 8.61 \\
157.9012909 & -39.9581795 & 15.93 & 151.41 & 14.94 & 23.00 & 11 & 67.2 & \phantom{1}0.250 & 0.008 & 0.033 & 0.40 & 8.73 & 8.70 & 8.63 \\
157.8947296 & -39.9578857 & 106.46 & 156.70 & 14.17 & 16.00 & 7 & 43.9 & \phantom{1}0.166 & 0.011 & 0.015 & 0.57 & 8.67 & 8.74 & 8.66 \\
157.8940125 & -39.9581223 & 116.35 & 152.45 & 16.12 & 17.74 & 8 & 46.6 & \phantom{1}0.193 & 0.011 & 0.017 & 0.36 & 8.71 & 8.72 & 8.69 \\
157.8930511 & -39.9580383 & 129.62 & 153.96 & 17.85 & 18.76 & 16 & 79.7 & \phantom{1}0.365 & 0.011 & 0.033 & 0.24 & 8.68 & 8.76 & 8.61 \\
157.8907776 & -39.9563713 & 160.98 & 183.96 & 20.72 & 21.08 & 29 & 120.0 & \phantom{1}0.656 & 0.013 & 0.052 & 0.46 & 8.67 & 8.74 & 8.61 \\
157.8907776 & -39.9568939 & 160.98 & 174.56 & 21.29 & 21.37 & 39 & 161.5 & \phantom{1}0.887 & 0.012 & 0.073 & 0.41 & 8.72 & 8.70 & 8.58 \\
157.8901825 & -39.9572258 & 169.20 & 168.58 & 23.25 & 23.28 & 40 & 184.2 & \phantom{1}0.920 & 0.014 & 0.068 & 0.50 & 8.71 & 8.66 & 8.54 \\
157.8900452 & -39.9567871 & 171.09 & 176.48 & 23.08 & 23.30 & 33 & 112.3 & \phantom{1}0.752 & 0.022 & 0.035 & 0.75 & 8.73 & 8.66 & 8.56 \\
157.8918915 & -39.9590683 & 145.62 & 135.42 & 22.70 & 24.16 & 47 & 188.8 & \phantom{1}1.060 & 0.017 & 0.063 & 0.84 & 8.74 & 8.66 & 8.61 \\
157.8919373 & -39.9594231 & 144.98 & 129.03 & 23.48 & 25.41 & 39 & 183.7 & \phantom{1}0.897 & 0.012 & 0.073 & 0.58 & 8.67 & 8.71 & 8.59 \\
157.8942871 & -39.9588661 & 112.56 & 139.06 & 17.71 & 20.70 & 9 & 62.1 & \phantom{1}0.200 & 0.009 & 0.023 & 0.37 & 8.64 & 8.70 & 8.64 \\
157.8962097 & -39.9587593 & 86.04 & 140.98 & 14.88 & 19.69 & 40 & 110.0 & \phantom{1}0.908 & 0.021 & 0.044 & 0.45 & 8.75 & 8.73 & 8.57 \\
157.8966827 & -39.9594955 & 79.51 & 127.73 & 17.07 & 23.69 & 56 & 160.6 & \phantom{1}1.274 & 0.015 & 0.090 & 0.38 & 8.77 & 8.69 & 8.58 \\
157.8959198 & -39.9601326 & 90.04 & 116.26 & 19.82 & 26.72 & 16 & 111.0 & \phantom{1}0.358 & 0.007 & 0.049 & 0.28 & 8.69 & 8.72 & 8.59 \\
157.8925171 & -39.9582062 & 136.98 & 150.94 & 19.38 & 20.20 & 30 & 127.5 & \phantom{1}0.694 & 0.011 & 0.060 & 0.25 & 8.70 & 8.76 & 8.62 \\
157.8982544 & -39.9577217 & 57.82 & 159.65 & 10.29 & 15.49 & 4 & 29.0 & \phantom{1}0.099 & 0.010 & 0.010 & 0.86 & 8.78 & $\cdots$ & 8.73 \\
157.8997650 & -39.9596252 & 36.99 & 125.39 & 17.78 & 27.62 & 8 & 92.7 & \phantom{1}0.190 & 0.005 & 0.038 & 0.57 & 8.75 & 8.68 & 8.61 \\
157.9006042 & -39.9586411 & 25.40 & 143.10 & 15.33 & 23.94 & 12 & 85.1 & \phantom{1}0.275 & 0.009 & 0.030 & 0.58 & 8.71 & 8.69 & 8.62 \\
157.8976135 & -39.9584541 & 66.67 & 146.47 & 12.96 & 18.78 & 7 & 45.0 & \phantom{1}0.163 & 0.009 & 0.017 & 0.55 & 8.64 & 8.72 & 8.64 \\
157.8942413 & -39.9597778 & 113.20 & 122.65 & 20.53 & 24.99 & 4 & 23.1 & \phantom{1}0.086 & 0.012 & 0.008 & 0.42 & 8.67 & 8.76 & 8.66 \\
157.8916016 & -39.9574585 & 149.62 & 164.39 & 20.03 & 20.09 & 40 & 131.5 & \phantom{1}0.912 & 0.014 & 0.067 & 0.28 & 8.70 & 8.73 & 8.56 \\
157.8937378 & -39.9590454 & 120.14 & 135.83 & 19.13 & 21.91 & 9 & 62.5 & \phantom{1}0.203 & 0.008 & 0.026 & 0.32 & 8.65 & 8.71 & 8.65 \\
157.8929443 & -39.9604263 & 131.09 & 110.97 & 24.43 & 28.70 & 3 & 35.7 & \phantom{1}0.079 & 0.005 & 0.015 & 0.39 & 8.59 & 8.72 & 8.65 \\
157.8932648 & -39.9594345 & 126.67 & 128.83 & 21.04 & 24.02 & 6 & 54.8 & \phantom{1}0.137 & 0.007 & 0.021 & 0.53 & 8.62 & 8.69 & 8.64 \\
157.8929138 & -39.9597778 & 131.51 & 122.65 & 22.61 & 25.81 & 5 & 55.1 & \phantom{1}0.105 & 0.006 & 0.017 & 0.59 & 8.63 & 8.62 & 8.66 \\
157.8952026 & -39.9605598 & 99.93 & 108.57 & 21.92 & 28.78 & 5 & 49.7 & \phantom{1}0.105 & 0.005 & 0.021 & 0.29 & 8.65 & 8.73 & 8.64 \\
157.8989258 & -39.9593887 & 48.56 & 129.65 & 16.46 & 25.18 & 4 & 36.7 & \phantom{1}0.088 & 0.007 & 0.013 & 0.49 & 8.64 & 8.71 & 8.68 \\
157.8952332 & -39.9592209 & 99.51 & 132.67 & 17.46 & 22.00 & 9 & 54.3 & \phantom{1}0.194 & 0.009 & 0.022 & 0.34 & 8.67 & 8.69 & 8.62 \\
157.8999634 & -39.9586525 & 34.25 & 142.90 & 14.64 & 22.86 & 11 & 68.0 & \phantom{1}0.242 & 0.009 & 0.026 & 0.52 & 8.66 & 8.74 & 8.63 \\
157.9015350 & -39.9562645 & 12.56 & 185.88 & 10.91 & 14.98 & 4 & 18.5 & \phantom{1}0.081 & 0.011 & 0.007 & 0.52 & 8.83 & $\cdots$ & 8.77 \\
157.8994751 & -39.9564056 & 40.98 & 183.34 & 6.83 & 10.54 & 7 & 30.9 & \phantom{1}0.164 & 0.016 & 0.010 & 0.63 & 8.86 & 8.79 & 8.71 \\
157.8965302 & -39.9604034 & 81.62 & 111.38 & 20.35 & 28.41 & 6 & 66.5 & \phantom{1}0.135 & 0.006 & 0.024 & 0.51 & 8.65 & 8.71 & 8.64 \\
157.8934937 & -39.9608765 & 123.51 & 102.87 & 24.98 & 30.55 & 3 & 43.4 & \phantom{1}0.074 & 0.004 & 0.019 & 0.37 & 8.65 & $\cdots$ & 8.65 \\
157.8941650 & -39.9608192 & 114.25 & 103.90 & 23.93 & 30.10 & 2 & 26.0 & \phantom{1}0.052 & 0.004 & 0.012 & 0.32 & 8.62 & 8.71 & 8.68 \\
157.8939362 & -39.9604645 & 117.41 & 110.29 & 23.09 & 28.42 & 3 & 19.3 & \phantom{1}0.058 & 0.006 & 0.010 & 0.23 & 8.74 & 8.75 & 8.75 \\
157.9013367 & -39.9590340 & 15.30 & 136.03 & 17.56 & 27.32 & 4 & 56.0 & \phantom{1}0.098 & 0.005 & 0.022 & 0.72 & 8.61 & 8.75 & 8.63 \\
157.8889618 & -39.9570351 & 186.04 & 172.02 & 26.20 & 26.46 & 5 & 56.0 & \phantom{1}0.104 & 0.004 & 0.024 & 0.32 & 8.56 & 8.73 & 8.61 \\
157.8884277 & -39.9573555 & 193.41 & 166.25 & 27.96 & 28.14 & 1 & 22.4 & \phantom{1}0.031 & 0.004 & 0.009 & 0.47 & 8.55 & $\cdots$ & 8.69 \\
157.8976288 & -39.9579697 & 66.46 & 155.19 & 11.22 & 16.22 & 5 & 2.3 & \phantom{1}0.108 & 0.050 & 0.003 & 0.39 & 8.82 & 8.70 & 9.04 \\
157.9013062 & -39.9574013 & 15.72 & 165.42 & 12.85 & 19.33 & 4 & 24.6 & \phantom{1}0.092 & 0.009 & 0.011 & 0.33 & 8.74 & 8.76 & 8.67 \\
157.9013062 & -39.9568329 & 15.72 & 175.65 & 11.49 & 16.75 & 2 & 12.3 & \phantom{1}0.050 & 0.009 & 0.005 & 0.75 & 8.77 & $\cdots$ & 8.80 \\
157.9001770 & -39.9553299 & 31.29 & 202.70 & 6.17 & 7.78 & 8 & 20.7 & \phantom{1}0.171 & 0.024 & 0.007 & 0.77 & 8.82 & 8.79 & 8.70 \\
157.8994293 & -39.9556847 & 41.61 & 196.32 & 4.86 & 7.07 & 6 & 15.1 & \phantom{1}0.143 & 0.025 & 0.006 & 0.75 & 8.84 & $\cdots$ & 8.69 \\
157.9002533 & -39.9547272 & 30.24 & 213.55 & 6.17 & 6.66 & 8 & 19.0 & \phantom{1}0.182 & 0.030 & 0.006 & 0.89 & 8.88 & $\cdots$ & 8.71 \\
157.8929443 & -39.9548340 & 131.09 & 211.64 & 14.02 & 15.68 & 14 & 52.4 & \phantom{1}0.318 & 0.019 & 0.017 & 0.54 & 8.75 & 8.82 & 8.66 \\
157.8922577 & -39.9574242 & 140.56 & 165.01 & 18.38 & 18.52 & 24 & 86.1 & \phantom{1}0.552 & 0.013 & 0.042 & 0.22 & 8.63 & 8.74 & 8.60 \\
157.8946533 & -39.9550934 & 107.51 & 206.97 & 9.34 & 10.04 & 7 & 18.5 & \phantom{1}0.157 & 0.031 & 0.005 & 1.13 & 8.84 & 8.77 & 8.70 \\
157.8951111 & -39.9548836 & 101.19 & 210.74 & 8.04 & 8.94 & 6 & 14.8 & \phantom{1}0.132 & 0.032 & 0.004 & 0.80 & 8.84 & $\cdots$ & 8.68 \\
157.8967743 & -39.9545021 & 78.24 & 217.61 & 3.70 & 4.85 & 9 & 6.7 & \phantom{1}0.213 & 0.116 & 0.002 & 1.57 & 9.03 & $\cdots$ & 8.82 \\
157.9007874 & -39.9540176 & 22.87 & 226.32 & 8.21 & 8.24 & 5 & 13.3 & \phantom{1}0.125 & 0.027 & 0.005 & 0.90 & 8.85 & $\cdots$ & 8.77 \\
157.9013367 & -39.9539719 & 15.29 & 227.15 & 9.69 & 9.75 & 5 & 17.3 & \phantom{1}0.112 & 0.020 & 0.006 & 1.04 & 8.78 & 8.72 & 8.71 \\
157.9012451 & -39.9545021 & 16.55 & 217.60 & 8.98 & 9.45 & 7 & 23.8 & \phantom{1}0.154 & 0.020 & 0.008 & 0.74 & 8.76 & 8.77 & 8.72 \\
157.8918304 & -39.9563370 & 146.46 & 184.58 & 17.89 & 18.08 & 18 & 88.2 & \phantom{1}0.400 & 0.013 & 0.030 & 0.42 & 8.67 & 8.76 & 8.64 \\
157.8919067 & -39.9541969 & 145.40 & 223.10 & 17.06 & 20.26 & 26 & 89.1 & \phantom{1}0.581 & 0.019 & 0.031 & 0.55 & 8.71 & 8.76 & 8.60 \\
157.8968658 & -39.9555206 & 76.98 & 199.27 & 3.97 & 4.10 & 13 & 13.3 & \phantom{1}0.288 & 0.068 & 0.004 & 0.65 & 8.93 & 8.81 & 8.65 \\
\hline
\multicolumn{6}{c}{Total / mean} & 2915 & 69.0 & 66.366 & 3.221 & 0.027 & 0.62 & 8.78 & 8.75 & 8.66 \\
\end{longtable}
The first row is the {\sn} position and the second is the centre of the galaxy. The last row shows the sum of the individual regions for extensive properties (H$\alpha$ flux and SFR) and the average for the intensive properties (equivalent width, extinction, and metallicities). (1) Right Ascension. (2) Declination. (3), (4) MUSE pixel position. (5) Distance from the galaxy centre. (6) Deprojected distance from the galaxy centre (Section~\ref{sec:optifs}). (7) H$\alpha$ flux in $10^{-16}\mbox{erg\,s}^{-1}\,\mbox{cm}^{-2}$. (8) H$\alpha$ equivalent width. (9) Star formation rate in $10^{-3}\msunyr$. (10) Stellar mass in $10^9\,\msun$. (11) Specific star formation rate ($\equiv\mbox{SFR}/\mstar$). (12) Extinction. (13) Metallicity based on [SII], [NII], and H$\alpha$ fluxes \citep{dopita16}. (14) Metallicity based on [OII], [NII], H$\alpha$, and H$\beta$ lines \citep{pettini04}. (15) Metallicity based on [NII] and H$\alpha$ lines \citep{pettini04}.

\end{document}